\documentclass[fleqn,usenatbib]{mnras}

\usepackage{newtxtext,newtxmath}

\usepackage[T1]{fontenc}

\DeclareRobustCommand{\VAN}[3]{#2}
\let\VANthebibliography\thebibliography
\def\thebibliography{\DeclareRobustCommand{\VAN}[3]{##3}\VANthebibliography}

\usepackage{graphicx}	
\usepackage{amsmath}	

\title[A radio flash following a neutron star merger]{A candidate coherent radio flash following a neutron star merger}

\author[A.~ Rowlinson et al.]{A.~Rowlinson$^{1,2}$\thanks{E-mail: b.a.rowlinson@uva.nl}, I.~de Ruiter$^{1}$, R.L.C.~Starling$^{3}$, K.M.~Rajwade$^{2}$, A.~Hennessy$^{3}$, R.A.M.J. Wijers$^{1}$, 
\and G.E.~Anderson$^{4}$, M.~Mevius$^{2}$, D.~Ruhe$^{1,5}$, K.~Gourdji$^{6,7}$, A.J.~van der Horst$^{8}$, S.~ter Veen$^{2}$, 
\and K.~Wiersema$^{9,10}$\\
$^{1}$Anton Pannekoek Institute for Astronomy, University of Amsterdam, Science Park 904, 1098 XH Amsterdam, The Netherlands\\
$^{2}$ASTRON, Netherlands Institute for Radio Astronomy, Postbus 2, 7990 AA, Dwingeloo, The Netherlands\\
$^{3}$School of Physics and Astronomy, University of Leicester, University Road, Leicester LE1 7RH, UK\\
$^{4}$International Centre for Radio Astronomy Research, Curtin University, GPO Box U1987, Perth, WA 6845, Australia\\
$^{5}$AI4Science (AMLab), Informatics Institute, University of Amsterdam, The Netherlands\\
$^{6}$Centre for Astrophysics and Supercomputing, Swinburne University of Technology, Hawthorn VIC 3122, Australia\\
$^{7}$OzGrav: ARC Centre of Excellence for Gravitational Wave Discovery, Hawthorn VIC 3122, Australia\\
$^{8}$Department of Physics, The George Washington University, 725 21st St NW, Washington, DC 20052, USA\\
$^{9}$Physics Department, Lancaster University, Lancaster LA1 4YB, UK\\
$^{10}$Centre for Astrophysics Research, University of Hertfordshire, Hatfield, AL10 9AB, UK\\
}

\date{Accepted XXX. Received YYY; in original form ZZZ}

\pubyear{2024}

\begin{document}
\label{firstpage}
\pagerange{\pageref{firstpage}--\pageref{lastpage}}
\maketitle

\begin{abstract}
In this paper, we present rapid follow-up observations of the short GRB 201006A, consistent with being a compact binary merger, using the LOw Frequency ARray (LOFAR). We have detected a candidate 5.6$\sigma$, short, coherent radio flash at 144 MHz at 76.6 mins post-GRB with a 3$\sigma$ duration of 38 seconds. This radio flash is 27 arcsec offset from the GRB location, which has a probability of being co-located with the GRB of $\sim$0.05\% (3.8$\sigma$) when accounting for measurement uncertainties. Despite the offset, we show that the probability of finding an unrelated transient within 40 arcsec of the GRB location is $<10^{-6}$ and conclude that this is a candidate radio counterpart to GRB 201006A. We performed image plane dedispersion and the radio flash is tentatively (2.4$\sigma$) shown to be highly dispersed, allowing a distance estimate, corresponding to a redshift of $0.58\pm0.06$. The corresponding luminosity of the event at this distance is $6.7^{+6.6}_{-4.4} \times 10^{32}$ erg s$^{-1}$ Hz$^{-1}$. If associated with GRB 201006A, this emission would indicate prolonged activity from the central engine that is consistent with being a newborn, supramassive, likely highly magnetised, millisecond spin neutron star (a magnetar).

\end{abstract}




\section{Introduction}

We do not know the maximum mass that a neutron star can have before it collapses to form a black hole. The detection of a 2 solar mass neutron star \citep{demorest2010} opened up the possibility that two typical 1.4 solar mass neutron stars could merge to form a supramassive neutron star instead of a black hole. This supramassive neutron star will be born rapidly rotating with strong magnetic fields and is referred to as a millisecond magnetar. Although supporting evidence for the millisecond magnetar model has been observed \citep{rowlinson2013,metzger2018,jordana2022}, the nature of the remnants of binary neutron star mergers remains highly debated. With its large rotational and magnetic energy reservoir, a millisecond magnetar is predicted to emit coherent radio emission within a few hours of formation \citep[e.g.][]{totani2013,zhang2014}, whereas black holes are not expected to produce coherent radio emission at this time unless they are actively accreting \citep{usov2000}. 

Binary neutron star mergers are typically detected via two key methods; short gamma-ray bursts \citep[GRBs;][]{abbott2017b} and gravitational wave events \citep{abbott2017}. Radio telescopes are used to conduct targeted searches for this short-lived coherent radio emission from GRBs and gravitational wave events. The emission we search for could occur very soon after or even during the merger so very low-latency observations are required \citep[for a detailed analysis of this see ][]{rowlinson2019b}. Due to the rapid response requirement, observations rely upon whole sky monitoring or very rapid repointing of the radio telescope. Early searches were thus either insufficiently deep \citep[e.g. ][]{balsano1998} or were not fast enough \citep[e.g. ][]{bannister2012}. With the advent of the next generations of sensitive low frequency radio telescopes that have no moving parts, are electronically steered and benefit from larger dispersion delays, such as the LOw Frequency ARray \citep[LOFAR; ][]{vanhaarlem2013} and the Murchison Widefield Array \citep[MWA; ][]{tingay2013}, we now have the capability to search for this emission. For the past few years, both LOFAR and the MWA have been triggering on GRBs, obtaining limits placing tight constraints on some of the theoretical models for this emission but no detections have been made to date \citep{kaplan2015,anderson2021,rowlinson2021,tian2022,curtin2022,hennessy2023}.

On 6th October 2020, the Neil Gehrels Swift Observatory \citep[hereafter {\it Swift} satellite;][]{gehrels2004} detected GRB 201006A, a short-hard GRB with an X-ray counterpart \citep{gropp2020}. No optical counterpart was detected, likely due to significant Galactic extinction along the line of sight (Galactic latitude of $b\sim9.7$ degrees). GRB 201006A does not have an identified host galaxy to deep limits in the near-infrared images at the location of the X-ray counterpart \citep{Fong22}. LOFAR automatically triggered observations of GRB 201006A, with a 2 hour imaging observation at 144 MHz starting 4.75 minutes following the GRB. 

In this paper, we present the LOFAR observations of GRB 201006A and the implications of these observations. In Section \ref{sec:Obs}, we describe the observations of this GRB and the processing strategies. In Section \ref{sec:transientHunt}, we outline the transient search strategy applied to these LOFAR data and the filtering results. Section \ref{sec:dedispersion} conducts image plane dedispersion on the transient candidate detected in the transient search. Finally, in Section \ref{sec:models}, we compare the observed radio flash to the theoretical model predictions for coherent radio emission from binary neutron star mergers.

Throughout this work, we adopt a cosmology with $H_0 = 71$ km s$^{-1}$ Mpc$^{-1}$, $\Omega_m = 0.27$ and $\Omega_{\Lambda}= 0.73$. Quoted errors are 1$\sigma$.

\section{Observations and analysis}
\label{sec:Obs}

\subsection{{\it Swift} detection and X-ray afterglow}
\label{sec:swiftObs}

The {\it Swift} Burst Alert Telescope \citep[BAT;][]{barthelmy2005} triggered and located GRB 201006A (trigger=998907) on 2020 October 6 at 01:17:52\,UT \citep{gropp2020}. {\it Swift} slewed immediately to the burst, and X-Ray Telescope (XRT) \citep{burrows2005} observations began 83.9 s after the BAT trigger, locating the X-ray afterglow to within a 90\% error region of 2.1$''$ radius at a position of RA: 61.89270 degrees, Dec: 65.16462 degrees (J2000) \citep{goad2020}. Data were processed at the UKSSDC \citep{Evans09}. GRB 201006A is classified as a short GRB with a $T_{90}$ duration of $0.49 \pm 0.09$ seconds (15--150 keV). High visual extinction due to the low Galactic latitude ($b\sim9.7$ degrees, $A_V\sim3.5$ mag \citep{Schlafly}) prevents deep host searches, while a K-band search has identified a faint extended source within the XRT error region though the probability of chance alignment is $>20\%$ \citep{Fong22}.

\begin{figure}
\centerline{\includegraphics[width=0.5\textwidth]{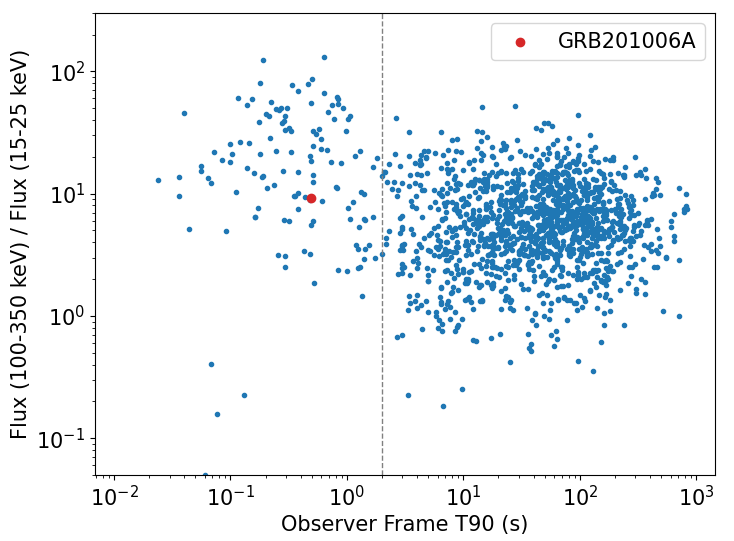}}
\caption{This plot shows the observed hardness ratio versus the T90 durations of GRBs detected by {\it Swift}. The dashed black line is the rough delineator between short and long GRBs at the T90 duration of 2 seconds. The location of GRB\,201006A (red point) is fully consistent with the short GRB population. 
\label{fig:hardness_duration}}
\end{figure}

\begin{figure}
\centerline{\includegraphics[width=0.5\textwidth]{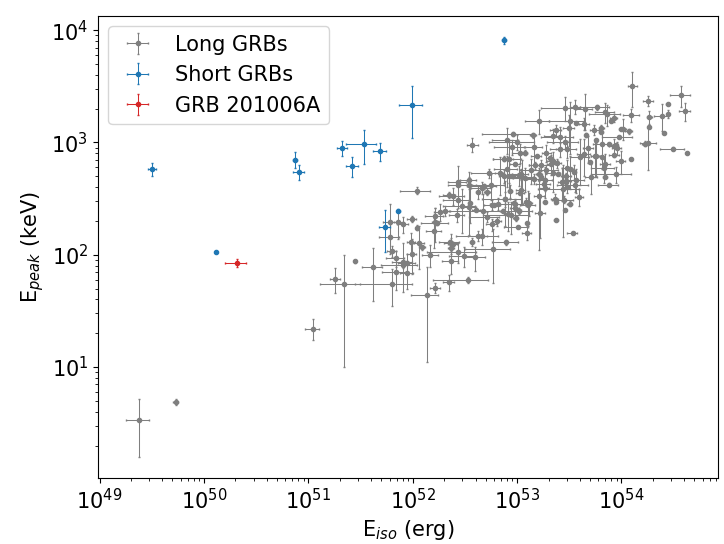}}
\caption{This plot shows the Amati relation for GRBs, which is an observed correlation between $E_{\rm peak}$ and $E_{\rm iso}$ \citep{Amati08}. Long and short GRBs follow different correlations \citep{Jia22,DAvanzo14}. The location of GRB\,201006A (red point) is fully consistent with the $E_{\rm peak} - E_{\rm iso}$ correlation for short GRBs (blue points).
\label{fig:amati_relation}}
\end{figure}

To further assess the nature of the GRB, we place it on both the hardness--duration diagram and the Amati relation, shown in Figures \ref{fig:hardness_duration} and \ref{fig:amati_relation}. GRBs show a correlation between peak energy, $E_{\rm peak}$, and isotropic energy, $E_{\rm iso}$ of the time-averaged prompt spectra \citep{Amati08} in which short GRBs are offset from the collapsar origin long bursts. In addition, the short and long populations show differing distributions in spectral hardness and duration \citep{Kouveliotou93}. The Amati relation is plotted for published long GRB \citep{Jia22} and short GRB \citep{DAvanzo14} samples, while the hardness--duration plot uses observed $\frac{50-100}{25-50}$ keV fluence and $T_{90}$ from the {\it Swift} BAT Catalogue \citep{Lien16}. Figures \ref{fig:hardness_duration} and \ref{fig:amati_relation} both show GRB 201006A (at $z=0.58\pm0.06$, see Section \ref{sec:distance}) to lie within the expected short GRB parameters for GRB samples with measured redshifts. 

\subsection{LOFAR triggered observations}
\label{sec:LOFARObs}

LOFAR observations of GRB 201006A were automatically triggered using the LOFAR Rapid Response Mode and starting at 2020 October 06 01:22:37 UTC, 4.75 minutes following the prompt gamma-ray emission. These observations comprised of 2 hours on the location provided by the gamma-ray observations, using the LOFAR High Band Antennas (HBAs) with a central frequency of 144 MHz and a bandwidth of 48 MHz (comprising 244 sub-bands each with a bandwidth of 195.3 kHz and a time resolution of 1 second). The observations used the Dutch array, comprising of 24 core stations and 14 remote stations. Following the observation of the GRB, a 10 minute calibrator observation was obtained using 3C147. A second observation was obtained 24 days later, starting at  2020 October 30 00:11:00 UTC, using the identical setup, less one core station, and with near-identical local sidereal time to the previous observation.

The LOFAR data were calibrated using the {\sc LINC} pipeline\footnote{\url{https://linc.readthedocs.io/en/latest/index.html}}, which was developed to automatically process these observations using standard LOFAR software and methods \citep{degasperin2019, vanweeren2016, williams2016, offringa2012, offringa2010}. The time resolution of 1 second was retained for the target visibility data. The calibrator and target visibility data were averaged in frequency to 48.82 kHz (4 channels per subband). The target subbands were then combined in groups of 10 and averaged in frequency to 97.64 kHz bins. As the target source is at the centre of the field, direction-dependent calibration was not required.

\begin{figure*}
\centerline{\includegraphics[width=\textwidth]{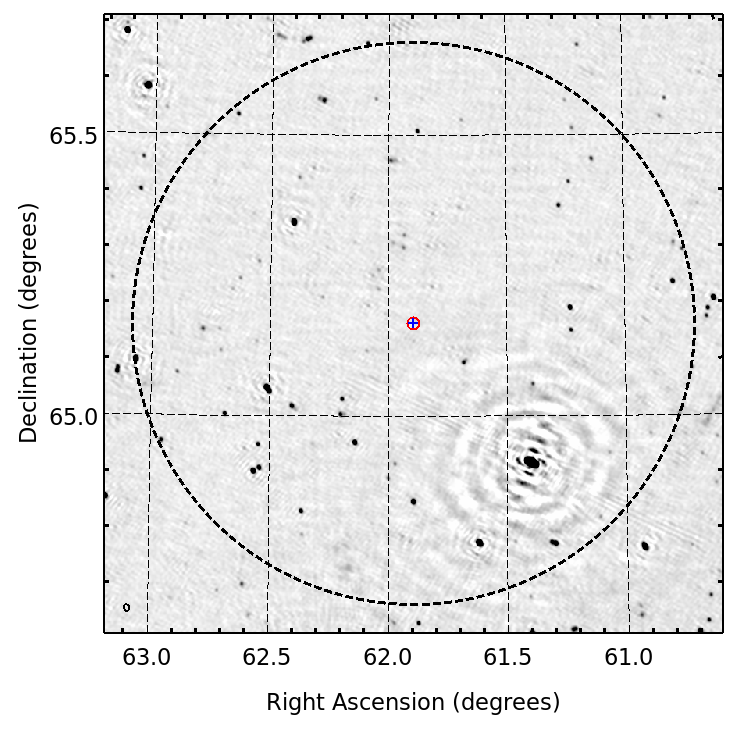}}
\caption{This is the full 2 hour image obtained by LOFAR at a central observing frequency of 144 MHz. The black dashed circle, with a radius of 0.5 degrees, shows the region searched for radio transient sources. The blue cross marks the X-ray position of GRB 201006A. The red circle corresponds to the 40 arcsec association radius centred on the X-ray position. The rms of the inner region of the image is 0.52 mJy. The restoring beam is illustrated in the lower left corner, with a major axis of 22 arcsec, a minor axis of 18 arcsec and a position angle of 20 degrees.
\label{fig:deepImg}}
\end{figure*}

The calibrated target data were imaged using {\sc WSClean} \citep{offringa2014} to create a deep image and a detailed sky model of the field. Standard imaging parameters were used in addition to a Briggs weighting (robustness of $-0.5$), a pixel scale of 5 arcsec, 4028$\times$4028 pixels, baselines up to 8 k$\lambda$ and a primary beam correction. Cleaning was conducted using an automatic threshold and up to 100,000 iterations. In Figure \ref{fig:deepImg}, we show the inner 1 degree $\times$ 1 degree area of the deep image of the field. The rms in this image is 0.52 mJy beam$^{-1}$ and the restoring beam is an ellipse with a major axis of 22 arcsec, a minor axis of 18 arcsec, with a position angle of 20 degrees. The sky model and corresponding model visibilities were output by {\sc WSClean}. No emission was detected at the location of GRB 201006A in the deep image, corresponding to a 3$\sigma$ upper limit of 1.6 mJy. We conduct a constrained fit, in which the source is forced to take the shape and orientation of the restoring beam, at the location of GRB 201006A giving a flux density of $0.8 \pm 0.8$ mJy.

\begin{figure}
\centerline{\includegraphics[width=0.5\textwidth]{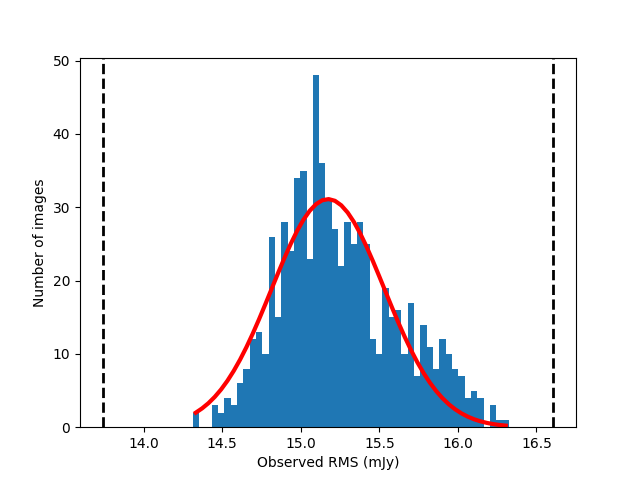}}
\centerline{\includegraphics[width=0.5\textwidth]{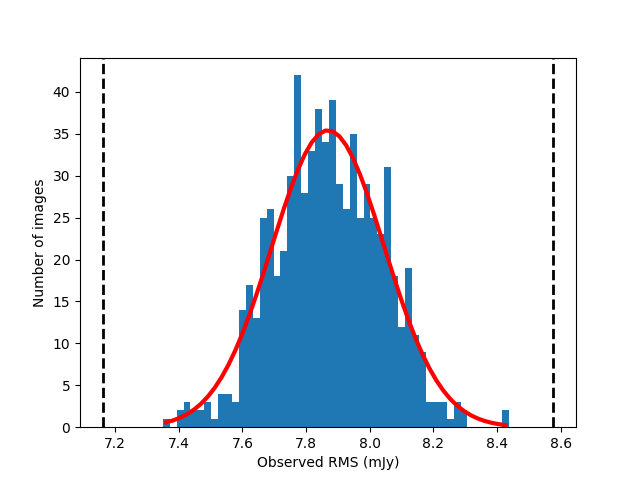}}
\caption{These histograms show the rms noise obtained in each of the snapshot images for the clean images with sources (top) and the dirty subtraction images (bottom). The black dashed lines show the rejection thresholds for low quality images. The typical rms of the clean images is $15.2\pm0.4$ mJy beam$^{-1}$ and the typical rms of the subtraction images is $7.9\pm0.2$ mJy beam$^{-1}$.
\label{fig:noiseHist}}
\end{figure}

We created two sets of time-sliced images with integration times of 10 seconds: cleaned images with all sources and a primary beam correction, and subtraction images with all the sources subtracted, no cleaning and with no primary beam correction.  We refer to the latter as dirty subtracted snapshot images. To create time-sliced dirty subtraction images, we first subtract the model visibilities (down to a 3$\sigma$ flux density threshold of 3 mJy to a radius of 2.8 degrees, obtained by {\sc WSClean}) from the calibrated data visibilities to obtain the subtracted visibilities \citep[this method has been developed and tested by ][]{fijma2023,deruiter}. The full 2 hours of subtracted visibilities were then imaged in 10 second snapshots using the full 48 MHz bandwidth and a pixel scale of 5 arcsec. The resulting subtraction images thus only contain subtraction artefacts and transient candidates. We note there is no detected source close to the GRB location and, hence, we do not expect any subtraction artefacts in the counterpart search area. For both datasets, we plot a histogram of the rms values from all of the images (determined using the inner 1/8th of the image), shown in Figure \ref{fig:noiseHist}, to determine the typical noise properties of the datasets and to check for low quality images. The typical rms noise in the cleaned images with all the sources is $15.2\pm0.4$ mJy beam$^{-1}$ and the typical rms noise in the subtraction images is $7.9\pm0.2$ mJy beam$^{-1}$, demonstrating a factor 2 improvement in detection sensitivity by using the subtraction images. 

\section{Radio transients search}
\label{sec:transientHunt}

In this section, we describe the transient search we conducted to determine if there is a radio source associated with GRB 201006A within the snapshot subtraction images. We outline the motivation for our source association radius and the detection threshold used to search for new sources.

\subsection{Source association radius}
\label{sec:sourceAssoc}

\begin{figure}
\centerline{\includegraphics[width=0.5\textwidth]{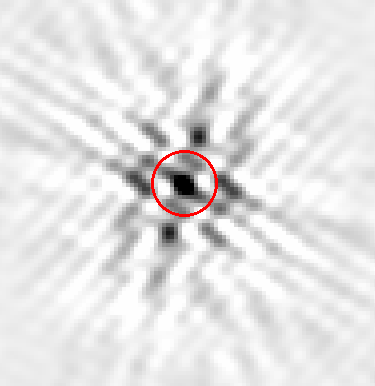}}
\caption{This figure shows the psf of the LOFAR dirty beam as calculated by {\sc WSClean} for one of the snapshot images. The red circle, centred on the psf and out to within the radius of the first sidelobe, has a radius of 40 arcsec and is defined to be the search radius for counterparts to GRB 201006A.
\label{fig:psf}}
\end{figure}

Radio sources can appear to move slightly in images due to measurement errors near to the rms noise, ionospheric effects and small calibration errors. To account for this in our search for counterparts to GRB 201006A, we need to determine the maximum radius out to which we would consider two sources to be potentially associated. To do this, we use the point spread function (psf) of the dirty beam, as calculated by {\sc WSClean}, for one of the snapshot images. The psf for one snapshot image is shown in Figure \ref{fig:psf}. We define the maximum search radius as being just within the first bright sidelobe of the dirty beam, corresponding to a radius of 40 arcsec. We do this, as the dirty beam shape is not deconvolved in the subtraction images, so we might expect it to affect the image within 40 arcsec of the true source location.

\subsection{Optimal detection threshold}
\label{sec:detectThresh}

\begin{figure}
\centerline{\includegraphics[width=0.5\textwidth]{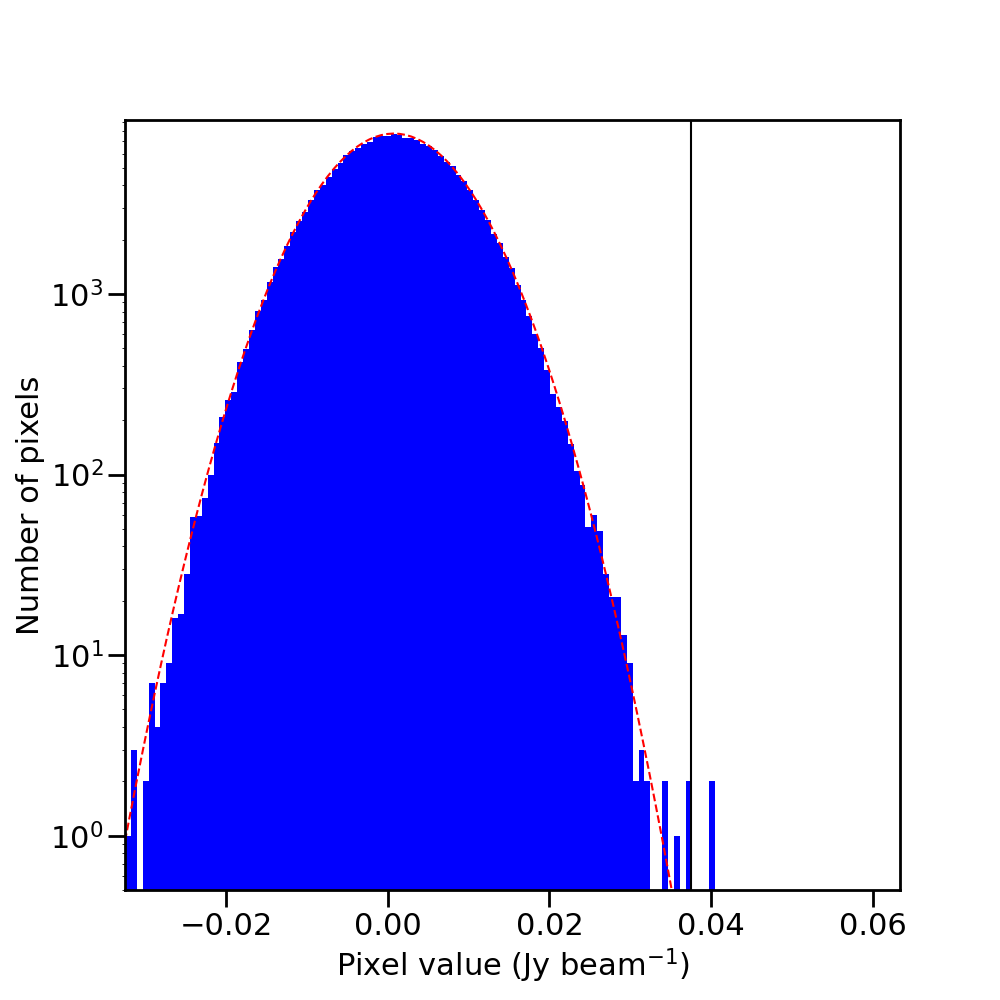}}
\caption{This is a histogram of all the pixel values within a radius of 40 arcsec from the position of GRB 201006A in all of the snapshot images. The red dashed line shows the fit to a Gaussian distribution. The black solid line corresponds to the detection threshold (4.7$\sigma$) required to give a probability of finding a spurious source caused by noise fluctuations to be less than 1\%.
\label{fig:pixelVals}}
\end{figure}

The next step is to determine the optimum detection threshold to search for sources associated with GRB 201006A. All the pixel values are extracted within a 40 arcsec radius of the enhanced X-ray position of GRB 201006A and are plotted as a histogram and fitted with a Gaussian distribution (see Figure \ref{fig:pixelVals}), which can then be used to determine the detection threshold \citep{rowlinson2022}. In this analysis, we require that the probability of a spurious source caused by noise fluctuations is less than 1\% corresponding to a detection threshold of 4.7$\sigma$ (shown by the black line in Figure \ref{fig:pixelVals}). We repeat this analysis for six background regions near to the GRB location (listed in Table \ref{tab:backgrounds}). The background regions all give the same recommended detection threshold of 4.7$\sigma$. To be conservative, we round this up to a detection threshold of 5.0$\sigma$ in the following analysis.

\begin{table}
\begin{center}
\caption{The positions of the six background regions used in this paper to compare to the findings obtained for GRB 201006A.
\label{tab:backgrounds}
}
\begin{tabular}{lll}
\hline
Background Region & RA & Dec \\
                  & (degrees) & (degrees) \\
\hline
1 & 62.4643 & 65.1854 \\
2 & 61.9437 & 65.0231 \\
3 & 61.9600 & 65.2114 \\
4 & 62.6837 & 65.3477 \\
5 & 61.0295 & 65.5890 \\
6 & 62.4026 & 64.6739 \\
\hline
\end{tabular}
\end{center}
\end{table}

\subsection{Search and filtering strategy}
\label{sec:TraP}

To search for counterparts to GRB 201006A and to determine the properties of candidate transients in the subtraction images used, we use the LOFAR Transients Pipeline \citep[{\sc TraP};][]{swinbank2015} to search for sources with a detection threshold of 5.0$\sigma$ within a radius of 0.5 degrees of the position of GRB 201006A. Assuming Gaussian noise properties, we predict that we will find three $>5 \sigma$ sources within 0.5 degrees throughout all of the surveyed images.

Using the {\sc TraP} sourcefinder, {\sc PySE} \citep{carbone2018}, sources are identified by finding islands of pixels that lie $>5\sigma$ above local rms noise. The detection significance of the source, in sigma, is obtained by dividing the peak pixel value by the local rms noise. As transient sources are expected to be point sources, we can assume that the source will take the shape of the restoring beam for that image. We note that the sources are not visually point sources in the subtraction images due to the shape of the psf (shown in Figure \ref{fig:psf}), but this assumption is reasonable to first order as the central part of the psf is a point source. We fit a two dimensional Gaussian, with its shape forced to be the shape and orientation of the restoring beam, to the detected source. The Gaussian fit provides the final position, flux density and associated fitting errors.

\begin{figure}
\centerline{\includegraphics[width=0.5\textwidth]{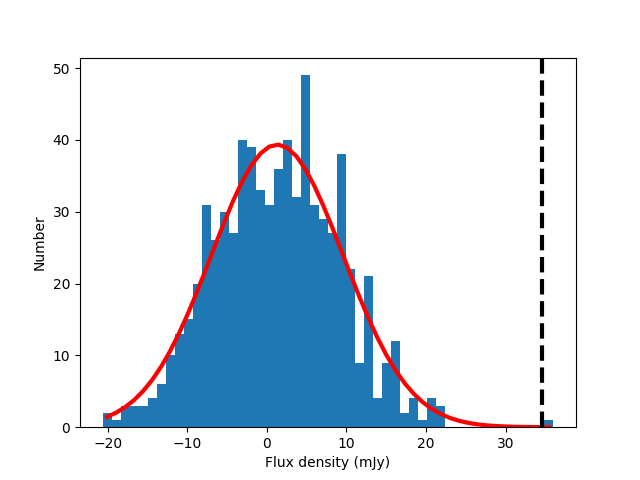}}
\centerline{\includegraphics[width=0.5\textwidth]{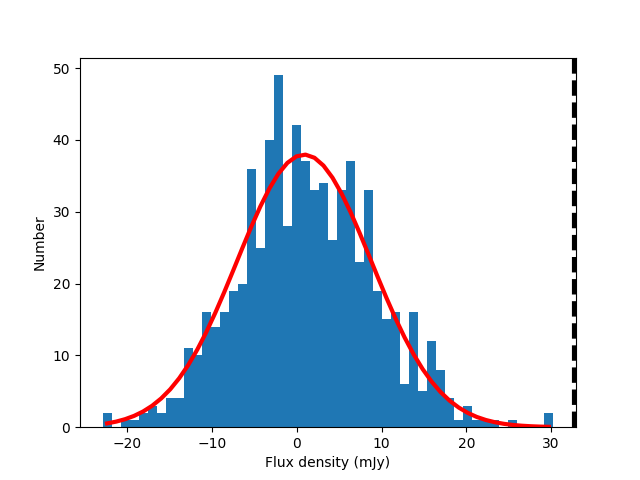}}
\caption{A forced flux density measurement was made for each transient source candidate for each snapshot image in the 2 hour observation. A histogram of the forced flux density measurements is then plotted for each individual source and fitted with a Gaussian distribution (shown by the red curve). The Gaussian distribution models the typical background noise at the location of the transient source candidate. If the flux density of the transient candidate is less than 4$\sigma$ deviant from the distribution (represented by the black dashed line), then we determine that the source is consistent with the noise properties at that location. Top: Transient candidate 6 from Table \ref{tab:candidates1}, which passed this test. Bottom: A random transient candidate that is rejected using this test. 
\label{fig:srcMonitoringHist}}
\end{figure}

\begin{figure*}
\includegraphics[width=0.48\textwidth]{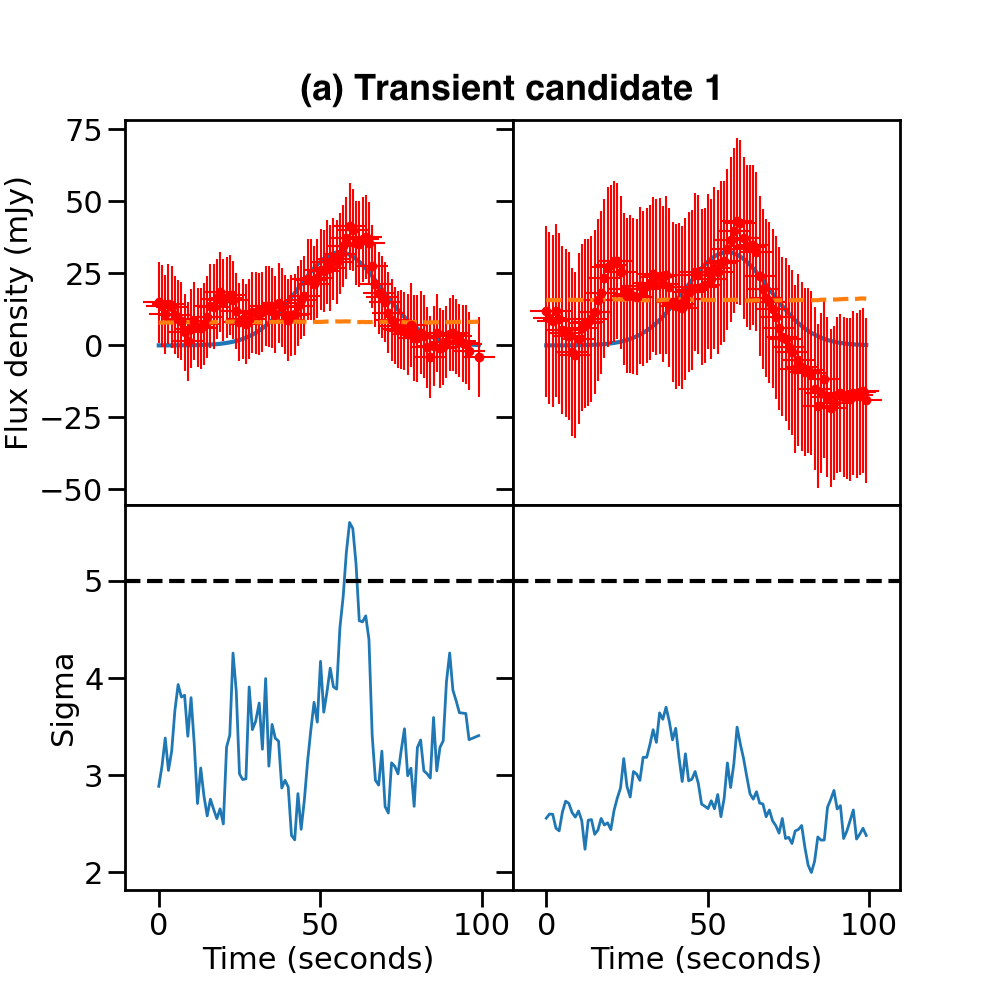}
\includegraphics[width=0.48\textwidth]{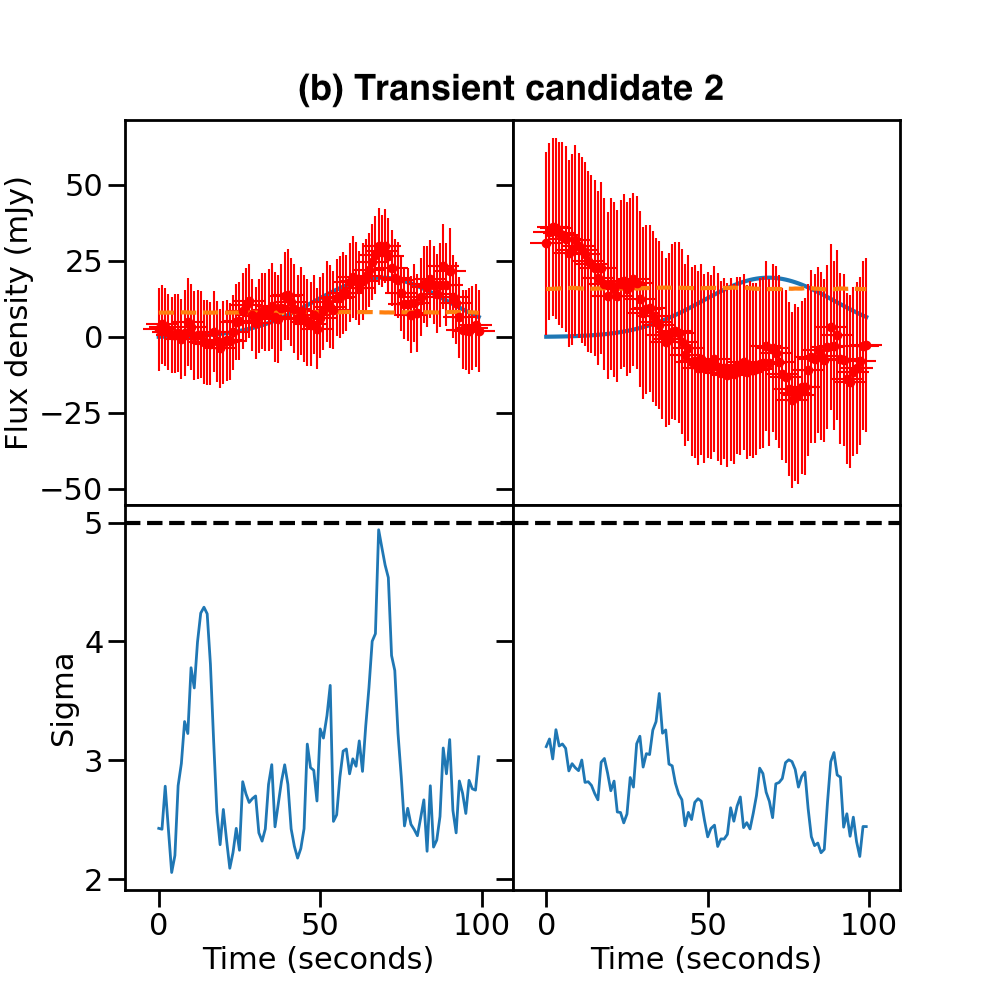}
\includegraphics[width=0.48\textwidth]{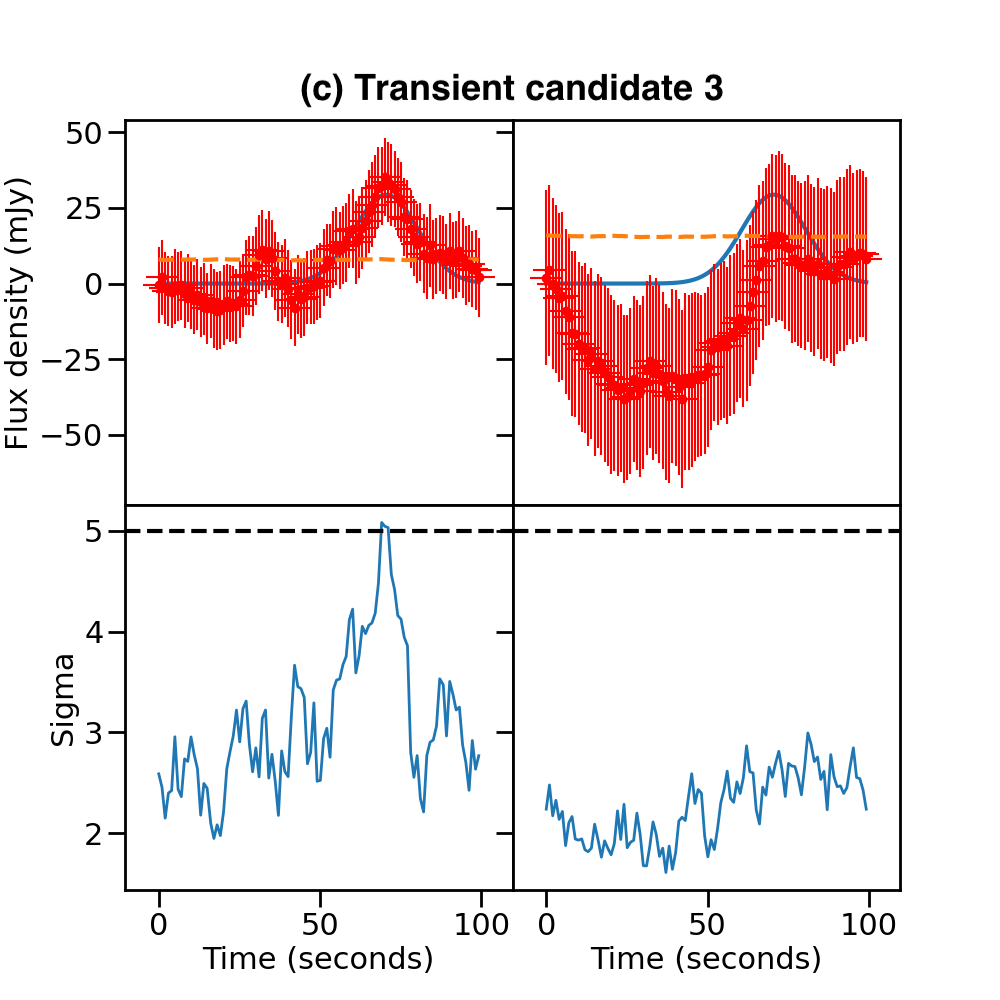}
\includegraphics[width=0.48\textwidth]{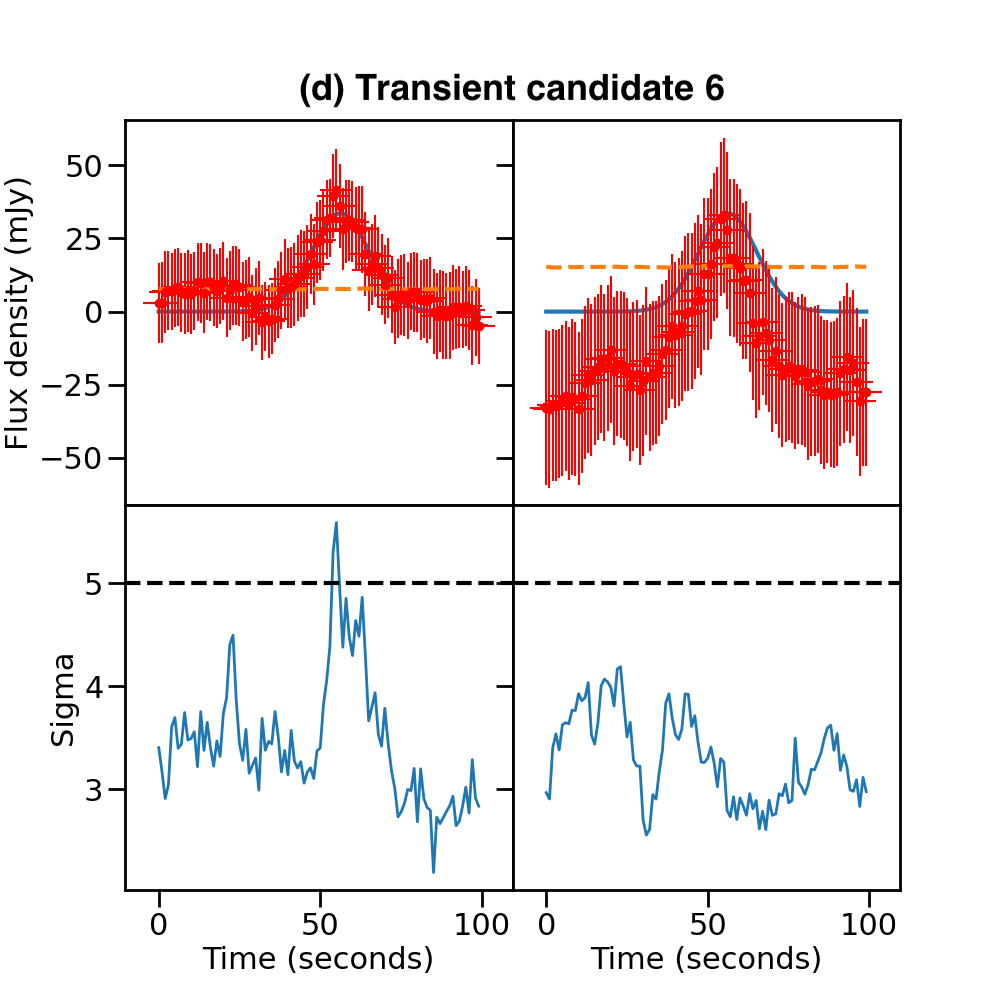}
\caption{The light curves of the four transient candidates given in Table \ref{tab:candidates2}. For each candidate, the plots show: Top row: The flux density light curve of the transient source created using 100 seconds of data and 10 sets of 10 second snapshots (each offset from the previous by 1 second) shown by the red data points. The zero time is arbitrarily chosen to ensure the full transient duration is covered by the light curve. The left column is created using the subtraction snapshot images, whereas the right column is created using the cleaned snapshot images containing all the sources. The solid blue line shows a Gaussian distribution fit to the data points from the subtraction snapshot images. The orange line shows the rms noise measured in each snapshot image. Bottom row: The detection significance of each measurement, with the black dashed line showing the 5$\sigma$ detection threshold.
In this plot, no dedispersion is applied (i.e. DM = 0 pc cm$^{-3}$).
\label{fig:Trans1}}
\end{figure*}

The {\sc TraP} outputs a list of newly identified sources and their flux densities for 10 images following detection. All sources detected are either subtraction artefacts or candidate transients. The initial candidate list contains 74 sources. We filter the list of candidates using the following strategies adapted from previous transient surveys \citep[such as ][]{rowlinson2022}:
\begin{enumerate}
    \item Remove sources close to the source extraction radius where the source finder is known to give artefacts. The number of remaining candidates dropped to 59 sources.
    \item Remove sources that are at the location of a source in the deep image (removal of subtraction artefacts). The number of remaining candidates dropped to 35 sources.
    \item Visual inspection of all candidates to remove sidelobes and subtraction artefacts of the bright, extended source 4C64.05 at RA: 61.4017 degrees Dec: 64.9181 degrees. All the artefacts associated with 4C64.05 were within 3 arcmin of its position. For reference, this is the bright source in the lower right of Figure \ref{fig:deepImg} with clear artefacts in the surrounding region. The number of remaining candidates dropped to 10 sources.
    \item Monitor the flux density at the location of each transient candidate through time to obtain a more accurate understanding of the local noise. A histogram of the measured flux densities was plotted for each source and fitted with a Gaussian distribution. Two example plots are shown in Figure \ref{fig:srcMonitoringHist}. Source candidates are rejected if the flux density is less than 4$\sigma$ deviant from the local noise measurements. The number of remaining candidates dropped to 6 sources and these candidates are listed in Table \ref{tab:candidates1}.  
    \item If the source is a noise artefact due to the telescope configuration, then it is likely to be present each time the position is observed at the same local sidereal time \citep{rowlinson2016}. Using the second observation on 2020 October 30, we are able to investigate the sources at the same local sidereal time. This second LOFAR observation was calibrated and imaged using the identical strategies as the first. Due to this observation only having near-identical local sidereal time to the original observation, one source does not have an image at the corresponding local sidereal time and is excluded from this analysis. We measure the flux density of each transient candidate in its corresponding local sidereal time image. The flux density measurements from the two observations are compared. If these flux density measurements are a near match (within measurement uncertainties) or the flux density is higher in the corresponding local sidereal time image, we determine that the transient candidate is likely a noise artefact due to the telescope configuration. 4 sources remained after this step, as shown in Table \ref{tab:candidates1}. 
    \item To constrain the peak detection significance, duration, peak time of the remaining transients, we re-imaged 10 second snapshot subtraction images for 100 seconds around each transient source. We created 10 batches of snapshot images, each offset in time by 1 second from the previous set, and used a constrained fit to obtain the flux density at the location of the transient source in each image. By combining the flux density measurements in a single figure, we are able to track the rise and fall of the transient source flux density. We then repeated this process with cleaned 10 second images containing all the sources to confirm if the source is consistent with being a sub-threshold source in those images. In Figure \ref{fig:Trans1}, we show the obtained light curves, as a function of both flux density and detection significance for the images with and without sources (i.e. DM = 0 pc cm$^{-3}$). The mid time of the transient is defined as the mean of the fitted Gaussian distribution with an additional 5 seconds to account for the midpoint of the snapshot (The time stamp in each 10 second snapshot corresponds the start time of that snapshot). If there is no evidence of the transient in the images containing sources or the light curve does not appear to follow a Gaussian shape (as expected for a dispersion smeared transient), we reject the transient candidate. We thus reject transient candidates 4 and 5. Two sources remained after this filtering step, as shown in Table \ref{tab:candidates2}.
\end{enumerate}

\begin{figure}
\centerline{\includegraphics[width=0.5\textwidth]{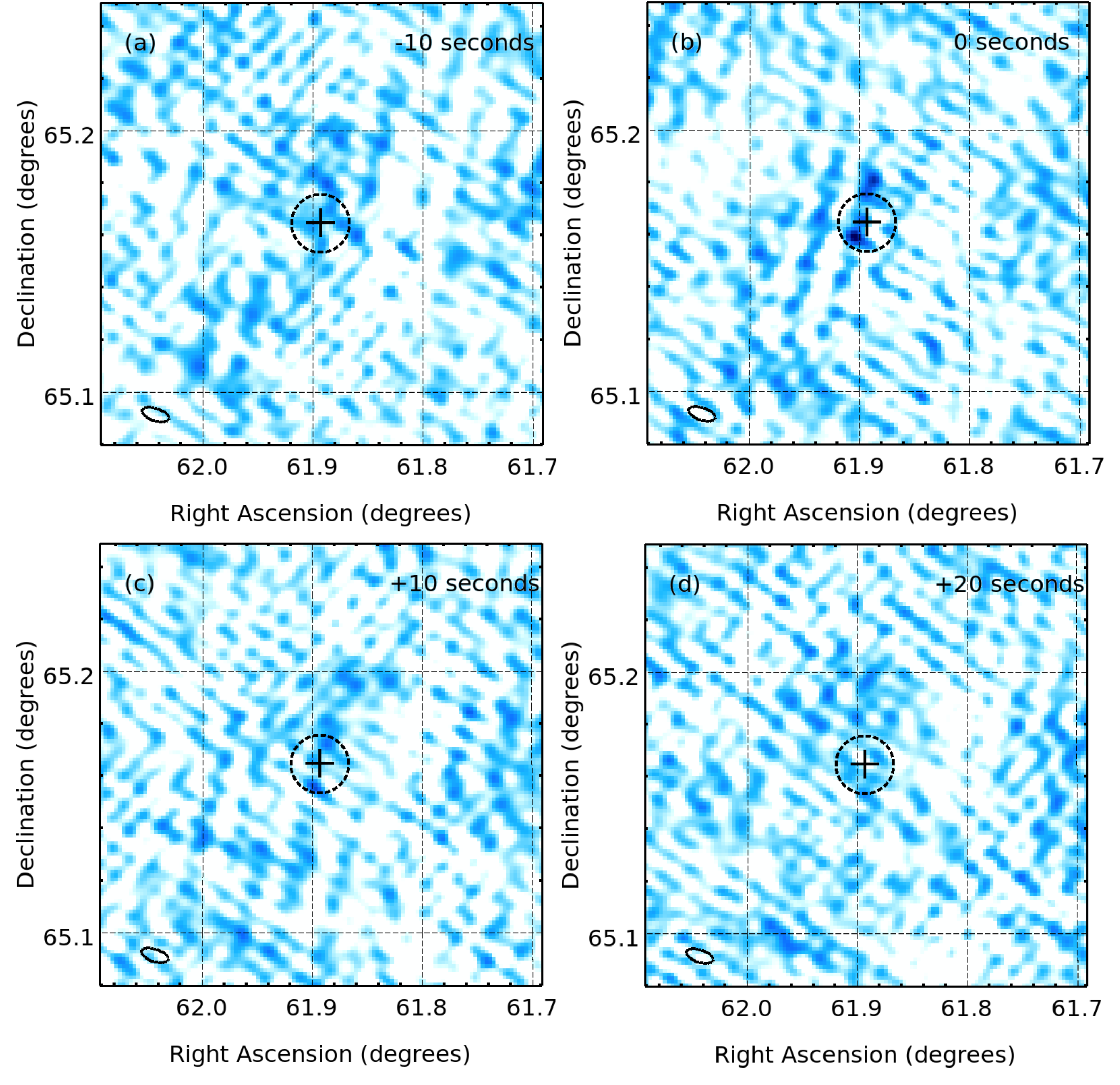}}
\caption{Subtraction images of the radio source location from 10 seconds prior to the emission and 20 seconds afterwards. The colour scale is matched between the figures, the black dashed line circles show the 40 arcsec counterpart search radius and the black cross marks the X-ray position of GRB 201006A. The restoring beam is shown in the bottom left corner. The 5.6$\sigma$ detection is within the search radius at time 0 seconds. The pixel scale is 5 arcsec with a restoring beam shape of major axis = 39 arcsec, minor axis = 17 arcsec and position angle of 162 degrees.
\label{fig:snapshots}}
\end{figure}

Of the remaining two transient candidates, candidate 1 lies 16 arcmin from the location of GRB 201006A and is unrelated to the GRB. Whereas candidate 6 lies within the 40 arcsec association radius for GRB 201006A. This transient is symmetrical in time and well fit by a Gaussian distribution, as expected for dispersion measure smearing. We find a peak S/N of 5.6$\sigma$ with a mid time of 2020 October 06 02:34:29.6 UTC, corresponding to 76.6 minutes after the GRB. As shown in Figure \ref{fig:Trans1}(d), in the cleaned images containing all sources, a constrained fit at the time and location of the transient source shows a comparable, but not significant, peak at the same time as in the subtraction images with a flux density consistent with the flux density of the observed transient source. The 3$\sigma$ duration of the radio transient source is 38 seconds. We note that this peak time is centred near the 10 second bins initially used. If the snapshot bins had been offset from this by more than a few seconds, we likely would not have detected this transient. Additionally, this source comprises all 4 pixels on or above the detection threshold at the GRB search region shown in Figure \ref{fig:pixelVals}. We find the position of this transient source in the peak detection image to be RA: 61.9042 degrees, Dec: 65.1588 degrees (J2000), with positional uncertainties of $\pm$(5.0,3.9) arcsec, and a flux density of $47\pm14$ mJy. We note this position is offset from the X-ray position of GRB 201006A by 27 arcsec. The subtraction images of this source are shown in Figure \ref{fig:snapshots}, along with 1 snapshot subtraction image prior to the detection and 2 snapshot subtraction images following. 

We trialled a wide range of imaging strategies to confirm the point source including: different baseline lengths, different image weighting, and different pixel scales. The point source remained observable in all the images produced. In addition, we conducted imaging at 5 seconds and the transient source was marginally detected in the two snapshots corresponding to the 10 second snapshot it was detected in. Further time and frequency slicing led to images where the source was below the detection threshold, likely due to higher noise levels. 

\begin{table*}
\begin{center}
\caption{The six remaining transient candidates following filtering step 4.
\label{tab:candidates1}
}
\begin{tabular}{llllllll}
\hline
Candidate & Time & RA & Dec & Detection S/N & Flux density & Flux density & LST Filter \\
  &  &  &  &  &  & LST matched Image & \\
  & (UTC) & (degrees) & (degrees) & (sigma) & (mJy) & (mJy) & ($\checkmark$/\texttimes)\\
\hline
1 & 01:32:43.3 & 62.0917 & 65.4233 & 5.4 & 41.9$\pm$14.1 & N/A & $\checkmark$ \\
2 & 01:45:42.4 & 61.3250 & 65.4071 & 5.0 & 36.6$\pm$12.6 & -0.5$\pm$14.0 & $\checkmark$ \\
3 & 01:48:44.6 & 61.2508 & 65.4911 & 5.1 & 38.9$\pm$12.8 & 5.2$\pm$15.5 & $\checkmark$ \\
4 & 01:49:54.7 & 61.6025 & 65.1851 & 5.2 & 43.2$\pm$13.1 & 16.3$\pm$16.2 & \texttimes \\
5 & 02:06:56.1 & 62.6521 & 65.3572 & 5.4 & 41.7$\pm$13.4 & 18.1$\pm$15.7 & \texttimes \\
6 & 02:34:24.4 & 61.9029 & 65.1591 & 5.0 & 44.8$\pm$14.4 & 4.0$\pm$16.3 & $\checkmark$ \\
\hline
\end{tabular}
\end{center}
\end{table*}

\begin{table*}
\begin{center}
\caption{The four remaining transient candidates after filtering step 5.
\label{tab:candidates2}}
\begin{tabular}{lllllll}
\hline
Candidate & Peak Time & Peak Detection & \multicolumn{2}{l}{Position at peak} & Light curve & Figure \\
  &  & S/N & RA & Dec &  & \\
  & (UTC) & (sigma) & (degrees) & (degrees) & (\checkmark/\texttimes) & \\
\hline
1 & 01:32:44.3 & 5.6 & 62.0914 & 65.4233 & \checkmark & \ref{fig:Trans1}(a)\\ 
2 & 01:45:48.1 & 4.9 & 61.3247 & 65.4071 & \texttimes & \ref{fig:Trans1}(b) \\ 
3 & 01:48:50.0 & 5.1 & 61.2508 & 65.4907 & \texttimes & \ref{fig:Trans1}(c) \\ 
6 & 02:33:29.6 & 5.6 & 61.9042 & 65.1588 & \checkmark & \ref{fig:Trans1}(d) \\ 
\hline
\end{tabular}
\end{center}
\end{table*}

\subsection{Consideration of position offset}

The detected transient source is offset from the GRB X-ray position by 27 arcsec. We showed in the previous section that this source passes a wide range of imaging tests, showing it is highly likely to be a real source. In this section, we consider if the position offset of the transient is consistent with expectations from radio images of this type from this dataset. First, we calculate the significance of the offset. Secondly, we conduct two key tests; an  analysis of the position of all detectable sources both in our dataset and using simulated sources. Thirdly, we consider randomly searching for sources within 40 arcsec of random locations. Finally, we consider the chance of finding an unassociated transient source within 40 arcsec of the position of GRB 201006A.

The 1$\sigma$ position uncertainty of the transient source is comprised of 3 different components: the statistical error on the position (calculated by the source finder), a systematic offset caused by the ionosphere and the position error on the X-ray source. The 1$\sigma$ position error on the RA and Dec measured for the transient source are $\pm(5.0,~3.9)$ arcsec respectively. The position shifts caused by the ionosphere during the observation is studied using 2 minute, cleaned snapshot images containing all the detectable sources in the field. The typical 1$\sigma$ position offset caused by the ionosphere during this observation is 3 arcsec. The 90\% enhanced X-ray position uncertainties for GRB 201006A is 2.1 arcsec \citep{goad2020}, corresponding to a 1$\sigma$ uncertainty of 0.8 arcsec. Combining these positional uncertainties in quadrature gives a total 1$\sigma$ position uncertainty of 7.1 arcsec. Thus, the observed 27 arcsec offset between the transient position and the X-ray position means that the GRB and radio source have a probability of being co-located of $\sim$0.05\% (3.8$\sigma$). This offset is notably large, but we can examine the offsets of sources in subtraction images of this kind to assess whether it is generally consistent.

To test the positional uncertainties of sources in the `dirty' source subtracted images used in this analysis, we split the problem into two key tests. Firstly, test the position uncertainty of sources in the `dirty' images and secondly test the position uncertainty in the subtraction images. 

\begin{figure}
\centerline{\includegraphics[width=0.5\textwidth]{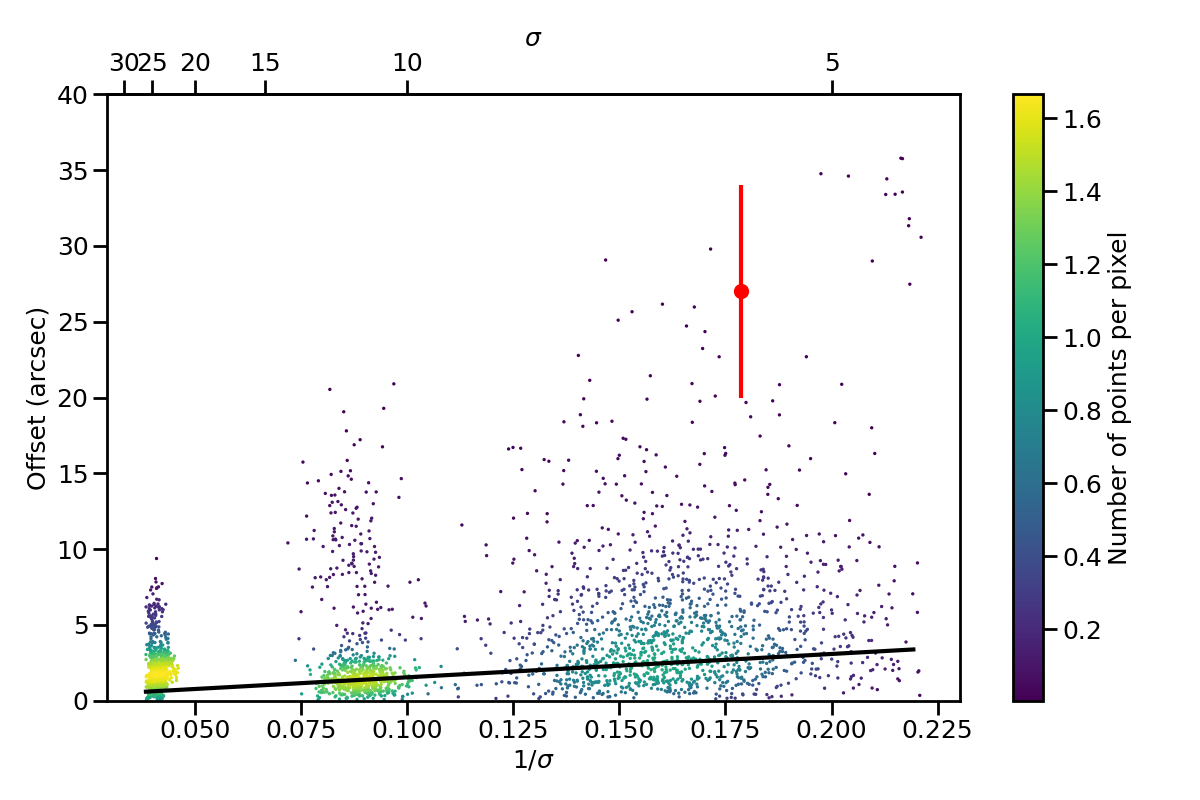}}
\centerline{\includegraphics[width=0.5\textwidth]{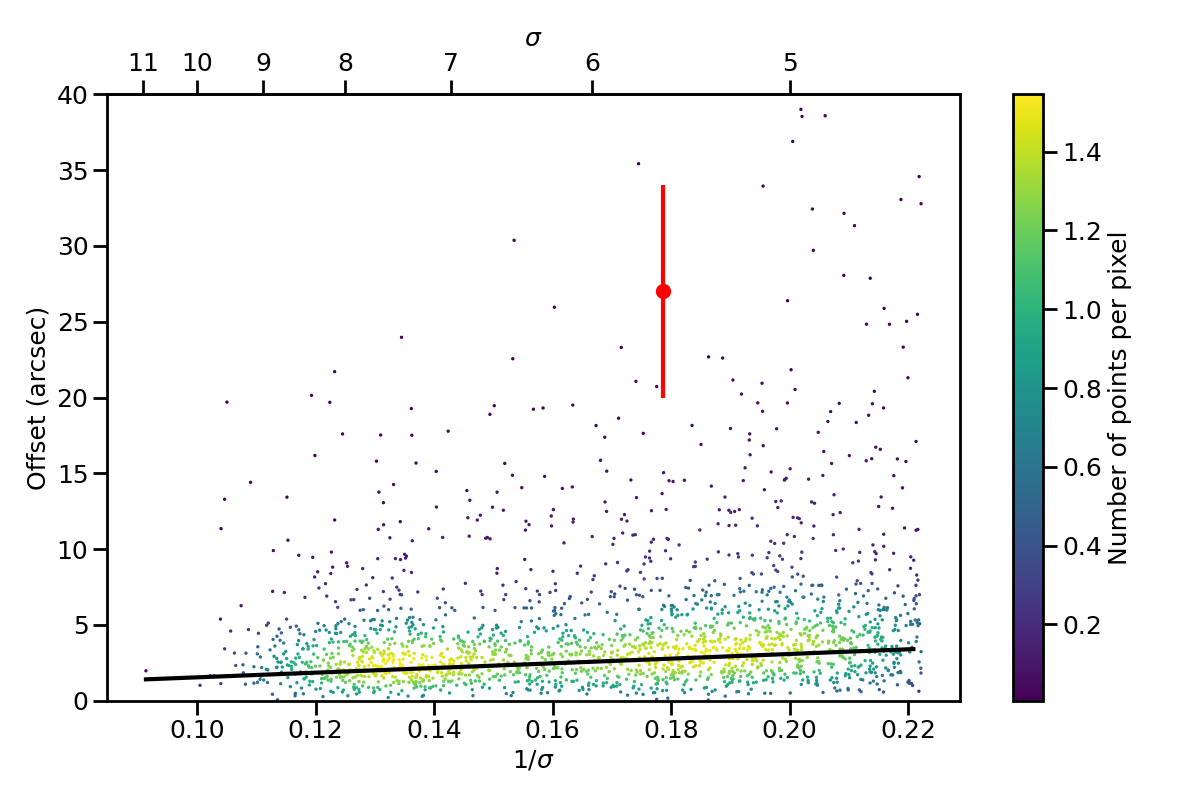}}
\caption{Top: This plot shows the offset of sources from their average position as a function of the signal to noise ratio of the source. The images used were `dirty' 10 second snapshot images containing all the sources in the field. The colour scale indicates the density of data points in a given pixel. The black solid line represents the predicted typical offset as a function of the signal to noise ratio \citep{loi2015}. The red datapoint shows the offset of the transient source from the GRB location, with the 1$\sigma$ positional uncertainty calculated for this source. Bottom: This plot instead shows the offsets of 8 simulated sources in the subtraction 10 second snapshot images.
\label{fig:offsets1}}
\end{figure}

To conduct the first test, we re-imaged the full dataset on 10 second timescales using the imaging settings to create the subtracted images but instead using the data containing all the sources - i.e. creating `dirty' and non-primary beam corrected 10 second snapshot images - giving a total of 712 images with $\sim$10 detectable sources per image given a 4.5$\sigma$ detection threshold. This test enables the study of positional offsets caused by the larger noise fluctuations observed in `dirty' images. However, the noise level is significantly higher than that observed in the subtraction images. We take the positions of the sources, as measured in the deep 2 hour image, and search for their associated counterparts, within the 40 arcsec search radius, in these snapshot images. In the case of multiple associations with a single source in the deep image (as can occur due to brighter sidelobes in `dirty' images), we only keep the source that is closest to the position in the deep image. In Figure \ref{fig:offsets1} (top), we then plot the offsets of the sources in these `dirty' images relative to the deep image positions as a function of the source signal-to-noise in the `dirty' image dataset. We also plot a black line showing the typical position measurement uncertainty for a source as a function of its signal to noise ratio ($\sigma$) \citep{condon1997,loi2015}. We note the majority of sources follow the expectation given their signal to noise ratio, but there is  scatter around this value. While it is still rare, at low signal to noise ratios the scatter extends to offsets of $\sim$35 arcsec. Given this distribution, the probability of a source having an offset of 27 arcsec or larger is 0.8\%. Thus, an offset of this size while unusual is not inconsistent with the observed offsets in the full dataset.

For the second test, we created 10 second subtraction images as for the original dataset but first remove 8 sources from the sky model that is subtracted. These 8 sources have flux densities comparable to the transient source (sources with flux densities of 30 -- 110 mJy in the deep, non-primary beam corrected, image), and will not be subtracted out, and thus remain visible in the 'subtracted' images. The resulting 712 `dirty', source subtracted images each have 8 simulated transient sources. These images are closest to our imaging scenario where there a transient source in the field remains after source subtraction as it is not present in the sky model. In Figure \ref{fig:offsets1} (bottom), we plot the offsets of these simulated sources relative to the positions of the 8 sources in the deep image as a function of the source signal-to-noise in the `dirty' image dataset. As with the previous test, the majority of the sources follow the expected trend as a function of their signal to noise ratio. However, there is a clear scatter around this correlation, which extends to the full search radius of 40 arcsec. Given this distribution, the probability of a source having an offset of 27 arcsec or larger is 0.5\%. Thus, an offset of this size is again unusual but not inconsistent with the observed offsets in the full dataset.

For the final test, we choose 250 random positions within a 0.5 degree radius of the position of GRB 201006A. These positions were chosen such that they were greater than 80 arcsec from the GRB position or from a source in the deep field with a flux density of $>50$ mJy, and were not close to the source extraction edge. Additionally, we exclude a 3 arcmin radius around 4C64.05 due to the known artefacts in this region. We also ensure the random positions are not at the same location by specifying that a random position is not closer than 80 arcsec to another random position. For each random position, we search for any sources detected at $\geq5\sigma$ within 40 arcsec of the position (the defined search radius) in all 712 images produced. Only 1 random position out of the 250 trialed had an association, giving a probability of finding a $\geq5\sigma$ source association by chance of 0.4 per cent. We note that the source associated with one of the positions was processed using our source filtering strategies in the previous section and was determined to be an artefact. Thus, we conclude that the chance association of the GRB position with a transient source within 40 arcsec is small.

Additionally, from transient surveys at low radio frequencies, we can calculate the probability of finding an unrelated transient of this brightness within the 40 arcsec radius from the X-ray position of GRB 201006A during our observation. Using images of 28 second duration from the Murchison Widefield Array at the higher frequency of 182 MHz, the transient surface density was constrained to be $<6.4\times10^{-7}$ deg$^{-2}$ at a sensitivity of 0.285 Jy \citep{rowlinson2016}. By assuming a flat spectrum and a cosmological population of sources (i.e. $N \propto S^{-\frac{3}{2}}$ where $N$ is the transient surface density and $S$ is the sensitivity of the observations), we can scale the transient surface density to the detected transient flux density of 47 mJy giving a transient surface density of $<1\times10^{-5}$ deg$^{-2}$ at 28 second timescales. We calculate the total sky area surveyed as the area of one image, $\left( \frac{40~{\rm arcsec}}{3600~{\rm arcsec}} \right)^{2}\pi =3.9\times10^{-4}$ deg$^2$, and multiply by the number of unique images in the dataset ($712-1$; accounting for one comparison image), giving a total sky area surveyed of $0.28$ deg$^{2}$. To compare to the 28 second survey conducted by MWA, we determine the unique sky area surveyed as being $\sim \frac{0.28}{3} = 0.09$ deg$^2$.  Thus we would expect to find $< 1\times10^{-5} \times 0.09 \sim 1\times10^{-6}$ transients with a duration of 28 seconds in our total surveyed sky area. Using 8 second snapshot images, at 144 MHz, created using the LOFAR Two Meter Sky Survey Data Release 1, the surface density of transients is $3.6 \times 10^{-8}$ deg$^{-2}$ at a sensitivity of 113 mJy \citep{deruiter}. Thus, in our total surveyed sky area at a sensitivity of 47 mJy, assuming the surface density is comparable for 8 and 10 second bins, we can show that we would expect to find $10^{-8}$ transients. Both of these surveys show that the chance of finding an unrelated radio transient by chance in the surveyed area of these images is negligible.

In conclusion, the offset between the transient source is notably larger than the typically observed offsets of sources in snapshot images from their positions in the deep image. However, there are sources, with comparable signal to noise ratios, that do have similar offsets in the two test snapshot image datasets created. Additionally, although a positional offset of this magnitude is unlikely, we have shown that the likelihood of an unassociated radio transient is orders of magnitude more unlikely. Given the presence of similarly offset sources on rare occasions in this dataset, the small chance alignment probability that a $\geq5\sigma$ source is near a random position, the confidence that the observed source is real and the extremely low probability that it is unrelated to the GRB, we determine this is most likely the radio counterpart to GRB 201006A.

\subsection{Consideration of second peak near transient candidate location}
\label{sec:artefact}

\begin{figure}
\centerline{\includegraphics[width=0.5\textwidth]{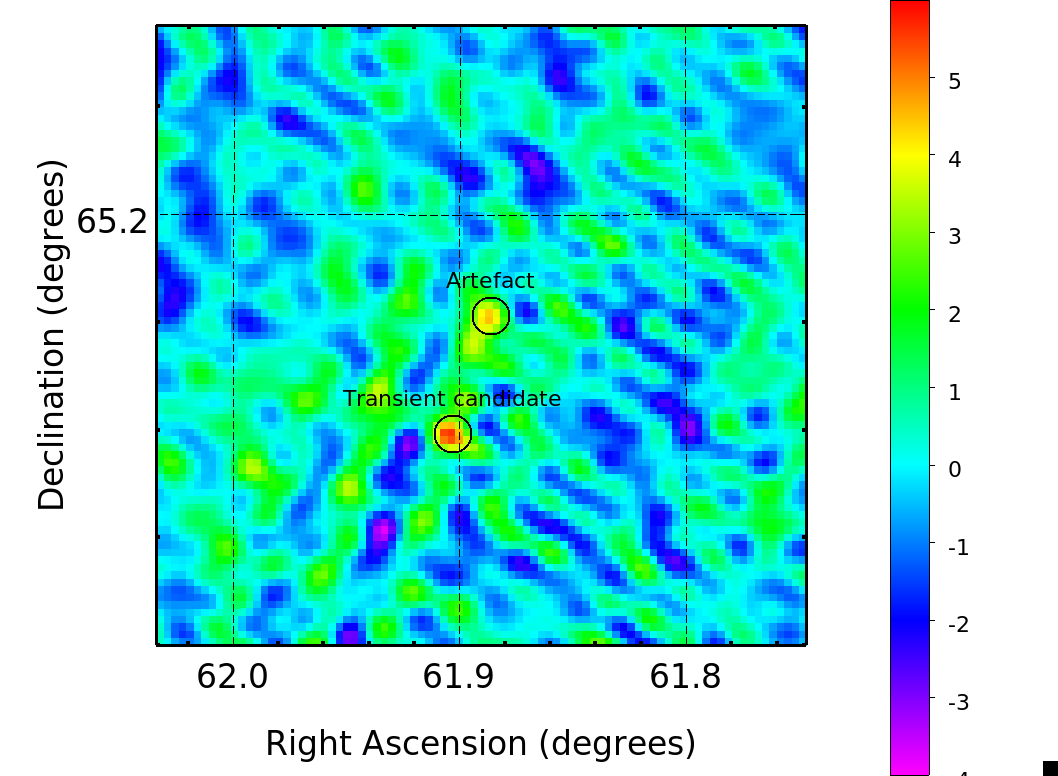}}
\caption{This figure shows the S/N map, produced by the {\sc PySE} source finder, of the image plotted in Figure \ref{fig:snapshots}b zooming in to the transient location. The colour scale is the S/N in $\sigma$. The transient candidate clearly stands out with a S/N of 5.6$\sigma$ whereas the northerly artefact is consistent with the noise.
\label{fig:artefact1}}
\end{figure}

We note that there is a second peak in the image 76 arcsec north from the transient candidate we associate with GRB201006A, as seen in Figure \ref{fig:snapshots}b with a position of RA: 61.8896 degrees, Dec: 65.1794 degrees (J2000). This second peak was not identified in the initial transient search using the 10 second snapshot images created for the transient search, as outlined in Section \ref{sec:LOFARObs}, as in all those images it only reaches a maximum S/N of 3.6$\sigma$, which is significantly lower than the optimal detection threshold of 4.7$\sigma$ and the conservative detection threshold we used of 5$\sigma$. Using a lower detection threshold on the image in Figure \ref{fig:snapshots}b (the peak brightness image for both the transient candidate and this second peak), we were able to extract the peak flux density of this source, giving $40.3\pm14.0$ mJy at a S/N of 4.5$\sigma$ (n.b. this is still lower than the optimal detection threshold). In Image \ref{fig:artefact1}, we show a signal-to-noise ratio map of the region containing the transient we associate with GRB 201006A and this second peak. The second peak has an order of magnitude lower S/N than the transient candidate, 4.5$\sigma$ and 5.6$\sigma$ respectively. Thus, given the 4.5$\sigma$ detection significance, the second peak is statistically most likely to be a noise fluctuation.

\begin{figure}
\centerline{\includegraphics[width=0.5\textwidth]{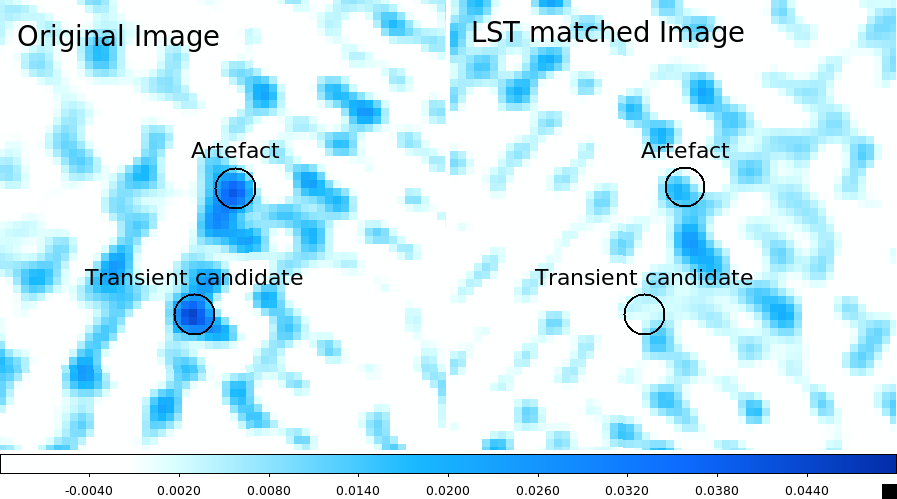}}
\caption{A zoom in on the location of the transient candidate and the northerly artefact in the detection image (left) and the LST matched image (right) with matching flux density scales.
\label{fig:artefact2}}
\end{figure}

We checked the location of this second peak in the LST matched image of the location, which was observed roughly one month following the triggered observation. In Figure \ref{fig:artefact2}, we show the detection image and the LST matched image. We note there is structured noise at the location of the second peak, which is not present for the transient candidate location. A forced fit at the location of this source gives a flux density measurement of $11.2\pm14.1$ mJy. We conclude that this second peak may be an artefact due to structured noise in the images. Whereas the lack of structured noise in the LST matched image at the transient candidate location is consistent with it being a real source in the detection image.

\begin{figure}
\centerline{\includegraphics[width=0.5\textwidth]{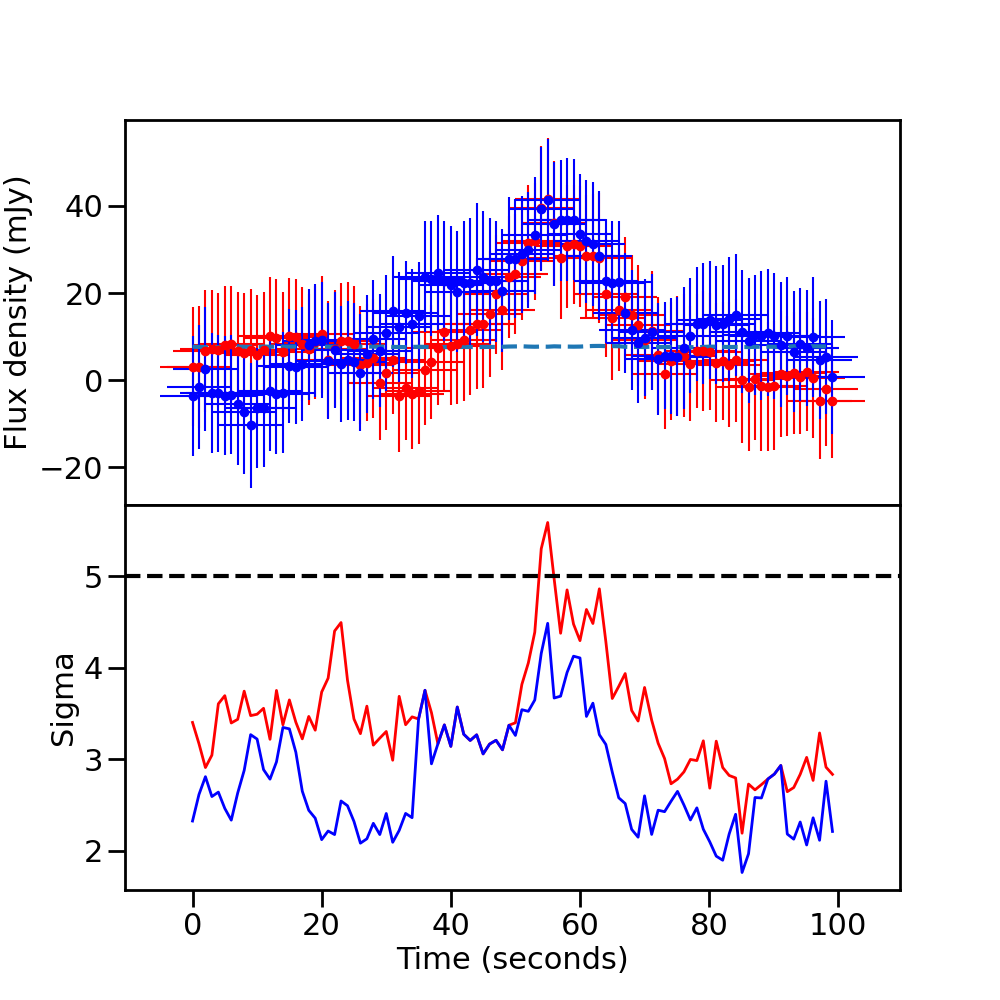}}
\caption{The light curves of the transient candidate (red) and the northerly artefact (blue) from 100 seconds of subtraction images (as Figure \ref{fig:Trans1}). Top: the flux density light curve and the typical noise in the images shown by the orange dashed line. Bottom: the detection significance of each measurement and the detection threshold shown by the black dashed line.
\label{fig:artefact3}}
\end{figure}

As a further test, we compare the light curve of the second peak to that of the transient candidate in the subtraction images. In Figure \ref{fig:artefact3}, we plot the observed light curve over the light curve of the transient candidate, showing both the flux densities and the S/N. We note that both sources peak at the same time and similar flux densities. The second peak has a broader light curve showing it is not following the identical behaviour as the transient candidate. 

In conclusion, the second peak is consistent with being a noise artefact. We note that the peak time and flux density of the artefact is similar to that of the transient candidate. However, the transient candidate is at a much higher detection significance and passes the filtering strategies showing it is consistent with a real source.

\section{Dedispersion analysis}
\label{sec:dedispersion}

If the observed radio emission is indeed originating from the extragalactic GRB 201006A, we expect its arrival time to be delayed due to its interaction with charged particles as it propagates from its origin to the radio telescope \citep{macquart2020}. The delay scales inverse quadratically with the frequency of the radio emission and gives a characteristic delay, quantified by a dispersion measure (DM). The dispersion measure can be quantified as:
\begin{equation}
{\rm DM} = \int_0^L n_{e}~dl\, \rm pc~cm^{-3},
\end{equation}
where $n_{e}$ is the electron density along the line of sight and $L$ is the distance to the source. For larger dispersion measures, the signal has propagated through more free electrons and hence is expected to be at a larger distance (assuming a constant density). If we know what the dispersion measure is, we can split the data into frequency and time slices, then recombine them correcting for the expected delay. This process is referred to as dedispersion. We do not know what the dispersion measure is for GRB 201006A a priori, so we trial different dispersion measure values to search for a dispersed signal using a few minutes of the data surrounding the detected radio source. The observed radio signal needs to be corrected for this dispersion in order to maximize its S/N. At the same time, it can be used to test whether the radio detection is truly astrophysical as only astrophysical signals will show a DM signature.

We created images spanning at least a few minutes before, during and after the transient event of varying integration/snapshot time steps ranging from 5 seconds to 20 seconds and split the frequency band into 16 frequency channels, each spanning 3~MHz. The coarser frequency resolution for the search was chosen in order to minimize loss of S/N when adding images at different frequencies rather than adding the visibilities at different frequencies (Tim Shimwell, Priv. Comm). Then, each dataset was corrected for the time delay due to interstellar dispersion using a custom made \textsc{C++} software suite \textsc{LORDS}\footnote{\url{https://git.astron.nl/rajwade/lordss}} \citep{rajwade}. For each set of images, we dedispersed the data over a DM range spanning from  0 to 1000~pc~cm$^{-3}$ with a step size of 3.9~pc~cm$^{-3}$. The only constraint to choose the granularity of the DM grid was that the delay across the LOFAR band due to the DM step size is less than the best resolution of the data (1 second). 

\begin{figure}
\centerline{\includegraphics[width=0.5\textwidth]{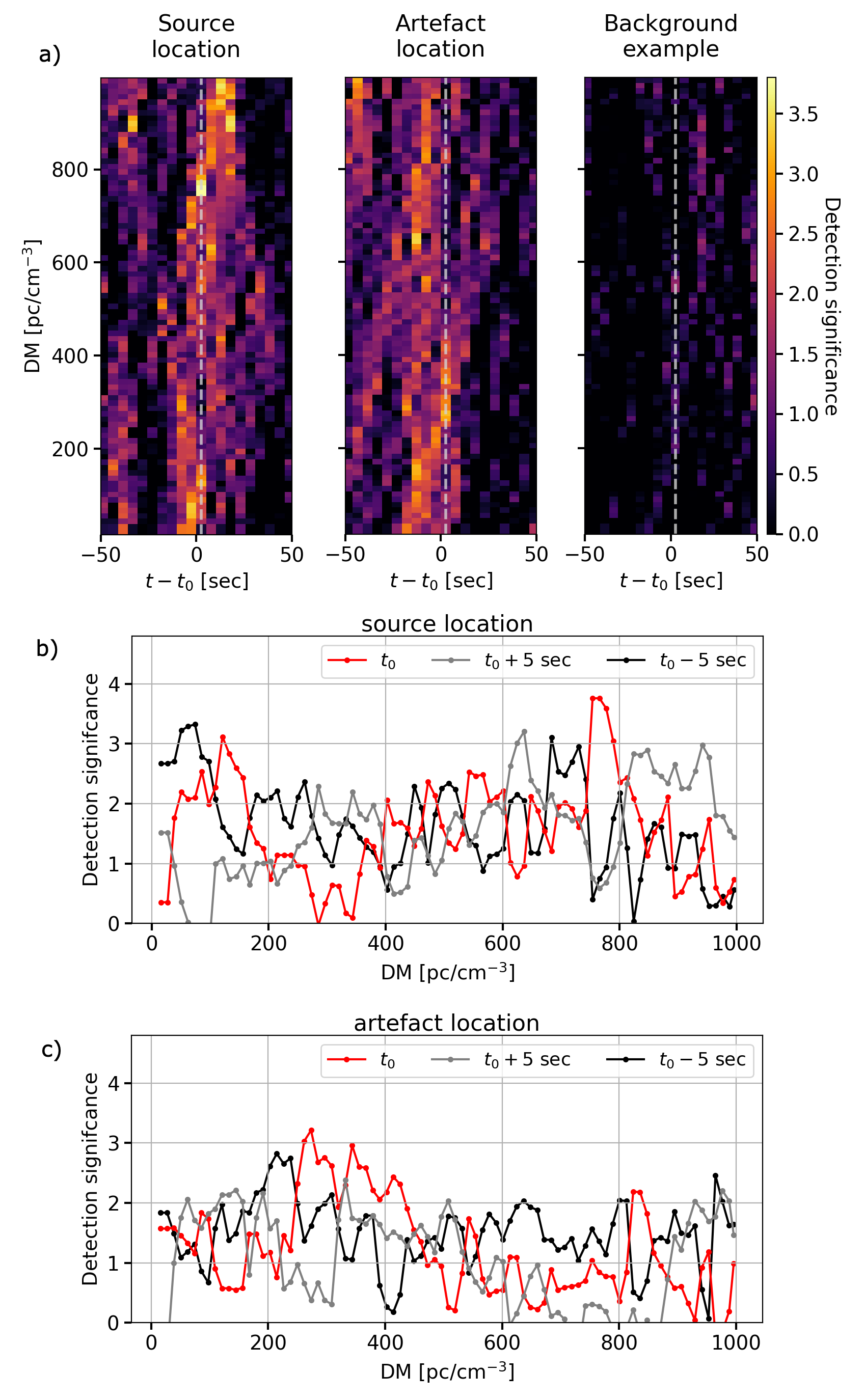}}
\caption{a) Detection significance at the location of the transient, location of the artefact, and a background pixel in the dedispersed images. The dedispersion process has been performed for each 5 second time bin (x-axis) over a range of DM trials (y-axis). The white dashed line shows the track along which we expect to see the dedispersed signal based on the initial estimate of the peak time of the signal. A 3.8$\sigma$ detection is visible at the source location at DM 740--800 pc cm$^{-3}$, compared to a 2.7$\sigma$ detection in the image where no dedispersion is performed (DM = 0 pc cm$^{-3}$ at $t_0-5$ sec). b) Vertical slices through the source location DM grid at $t_0$, $t_0+5$ sec and $t_0-5$ sec in grey, red and black respectively. c) Same as b) but for the artefact location.
\label{fig:dm_search}}
\end{figure}

An image timeseries was generated for each DM trial which was then run through a convolutional source finder \citep{ruhe2022detecting}. This source finder creates a detailed model of the noise in each image, which is necessary since the dedispersed images show noise structures that may not be accurately captured by traditional source finding methods. We give the source finder the pixel where the source is located and pick an additional 100 random pixels within a box of 500 by 500 (42 by 42 arcmin) pixels centered on the source location. The source finder then calculated the detection significance of each pixel at each timestep and each DM trial. As the transient source has a low S/N, we take the narrow DM step sizes and rebin them into blocks of three by taking the average of the three significance values. In Figure \ref{fig:dm_search}a we show the detection significance at the location of the transient, the location of the artefact and an example background pixel in the dedispersed images, for each time bin (x-axis) and a range of DM trials (y-axis). The white dashed line shows the track along which we expect to see the dedispersed signal based upon the initial estimate of the peak time of the signal. As the peak time is an estimate, we consider all signals at $\pm5$ seconds. We also take vertical slices through the data, for $t_0$, $t_0 + 5$ sec and $t_0 - 5$ sec, at the source location and the artefact location in Figures \ref{fig:dm_search}b and c respectively. At the source location a 3.8$\sigma$ detection is visible for DM values of 740--800 pc cm$^{-3}$ with a mid point of 770 pc cm$^{-3}$, compared to a 2.7$\sigma$ detection in the image where no dedispersion is performed (DM = 0 pc cm$^{-3}$ at $t_0 - 5$ sec). The range of DM values for the detection is likely caused by intra-channel smearing. 

We note that the dedispersed images are of a lower quality than the standard imaging technique, due to being composed of summed images rather than a single combined output from the imager, leading to the lower observed significance values. We also note that the source location is noisier than the background region, this is likely due to residual signal in the different bins being added together to create the dedispersed images. The background shows a more random uniform behavior with an average detection significance of 0$\sigma$. 

\begin{figure}
\centerline{\includegraphics[width=0.5\textwidth]{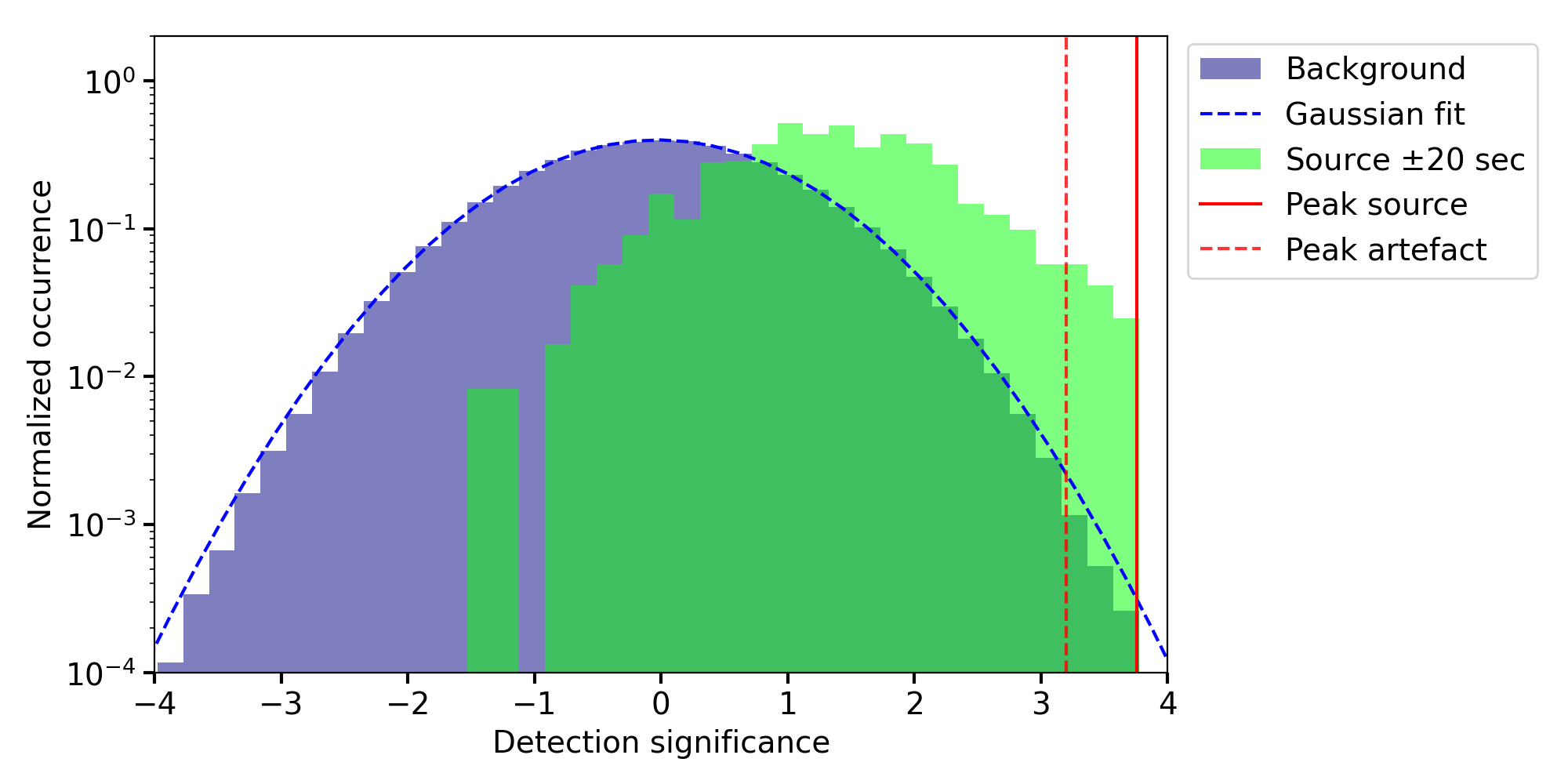}}
\caption{Distribution of detection significance values in 100 randomly selected background pixels (eg. right panel Figure \ref{fig:dm_search}a) in the dispersed images is shown by the blue histogram. The green histogram shows the same for the source location. We fit the background histogram with a Gaussian and show the peak detection significance at the source location, shown with a red vertical line, lies $3.9 \sigma$ away from the background distribution. For reference, the peak detection significance at the artefact location is shown with the red dashed vertical line ($3.2 \sigma$ away from the background distribution).
\label{fig:dispersion_hist}}
\end{figure}

In Figure \ref{fig:dispersion_hist} we show in blue the distribution of the detection significance values from the 100 randomly selected background pixels. The green histogram shows the values of the detection significance at the source location at a $t_0\pm 20$ second time window around our initial detection time. The source distribution is offset from the background and this is expected due to the presence of a source. The background signal follows a Gaussian distribution and we show that the peak detection significance at the source location is 3.9$\sigma$ from the background mean, which is consistent with the 3.8$\sigma$ detection (found in Figure \ref{fig:dm_search}). For reference, we show the peak detection significance of the artefact, which is 3.2$\sigma$ from the background mean.

A dispersion measure of 770 pc cm$^{-3}$ gives a dispersion delay of 109 seconds across the total LOFAR observing band (48 MHz bandwidth centred on 144 MHz), however the duration of the transient is $\sim$30 seconds in the DM 0 pc cm$^{-3}$ images. We note that many FRBs show significant frequency structures \citep{petroff2022} and the discrepancy for this radio source could be resolved if the signal has a frequency width of $\sim$12 MHz or if it has a steep spectral index across the band \citep{spitler2016}. We split the observations into a number of 12 MHz frequency bins (each offset by 2 MHz) and 30 second time bins (each offset by 2 seconds), assuming DM = 0 pc cm$^{-3}$, there is a hint that the radio source is indeed narrow band however we are unable to confirm this due to the low S/N of the radio source relative to background noise features.

In order to optimise the S/N of the transient, we attempted to estimate the most optimal width of the transient. To do that, we convolved the image timeseries for each DM trial with a Gaussian template with a range of widths (geometrically spaced from 5 to 31 seconds).  Each timeseries is median subtracted and then normalized such that the sum of the squares in the timeseries is unity. This is done to simplify the convolution without affecting the noise properties of the data. Then, we estimated the S/N for each width and time bin after convolving the timeseries with the set of Gaussian templates to make a S/N map in time and Gaussian filter width space for each DM trial. We find the intrinsic duration of the radio flare is consistent with being 7--8 seconds. We do note that while there is a marginal signal at the reported DM, we cannot conclusively show that the transient is dispersed, we also cannot rule it out completely and hence, use these estimates as nominal values for the paper.

Since we are estimating the optimal DM and width of the transient, its true significance is going to depend on the number of width and DM trials used in the search. In order to measure the true significance of the dispersed burst, we simulated 100 realizations of the DM timeseries grid (with the same DM range) with the same noise characteristics as the data. To do that we computed the mean and the standard deviation of each timeseries corresponding to each DM trial and generated 100 such timeseries with the same noise characteristics assuming standard Gaussian noise. These were then combined to make 86 DM trial grids. Then, we convolved these grids with a series of Gaussian filters with widths ranging from 5 to 31 seconds. We then generated a histogram of all the S/N pixels and computed the false alarm probability of the transient,
\begin{equation}
P_{\rm FAP} = P(S/N > S/N_{\rm transient}).
\end{equation}
We find a false alarm probability of 0.0046 for this radio flare which corresponds to a confidence of 2.4$\sigma$ that the observed dispersion is astrophysical in nature and hence we claim this to be a tentative detection of a dispersed signal.

\subsection{Consideration of second peak near transient candidate location}

In Section \ref{sec:artefact}, we noted a second peak north of the transient candidate and determined it was consistent with being a noise artefact and likely to be unrelated to the transient candidate. To confirm that the transient candidate does not demonstrate the same behaviour as the northern artefact, we repeat our analysis of the transient candidate for the location of the northerly artefact. 

In the second panel of Figure \ref{fig:dm_search}a, we plot the detection significance of the northerly artefact in the dedispersion images. The peak observed for the transient candidate at the DM range 740-800 pc cm$^{-3}$ is not present in the analysis for the northerly artefact. We find that the vertical slices through the DM grid at the artefact location in Figure \ref{fig:dm_search}c show no clear peak, compared to the vertical slices through the DM grid at the source location (Figure \ref{fig:dm_search}b). Even though the peak detection significance in the dispersed images at the artefact location is offset from the background detection significances, as indicated by the red dashed line compared to the blue histogram in Figure \ref{fig:dispersion_hist}, we conclude that there is no evidence that the artefact has a dispersion signature based on Figure \ref{fig:dm_search}c. A smoother and more gradual change in detection significance is expected around the appropriate DM trial for a truly dispersed signal. No astrophysical dispersion signature is expected for an imaging or calibration artefact.

\subsection{Distance constraint and comparison to known coherent radio transient population}
\label{sec:distance}

We calculated the expected Galactic and Galactic halo components of the dispersion measure along the line of sight to GRB 201006A. We find a Galactic component of 190 pc cm$^{-3}$ \citep{yao2017,yao2019} and a Galactic halo component of 38 pc cm$^{-3}$ \citep{yamasaki2020}, giving a total contribution of 228 pc cm$^{-3}$. As GRB 201006A is an extragalactic source, we expect to find a dispersion measure greater than this value. GRB 201006A is a short GRB which is likely offset from its host galaxy (it is identified as a ``hostless'' GRB \citep{Fong22}) and, thus, the excess dispersion measure is expected to be from the intergalactic medium.

\begin{figure*}
\centerline{\includegraphics[width=\textwidth]{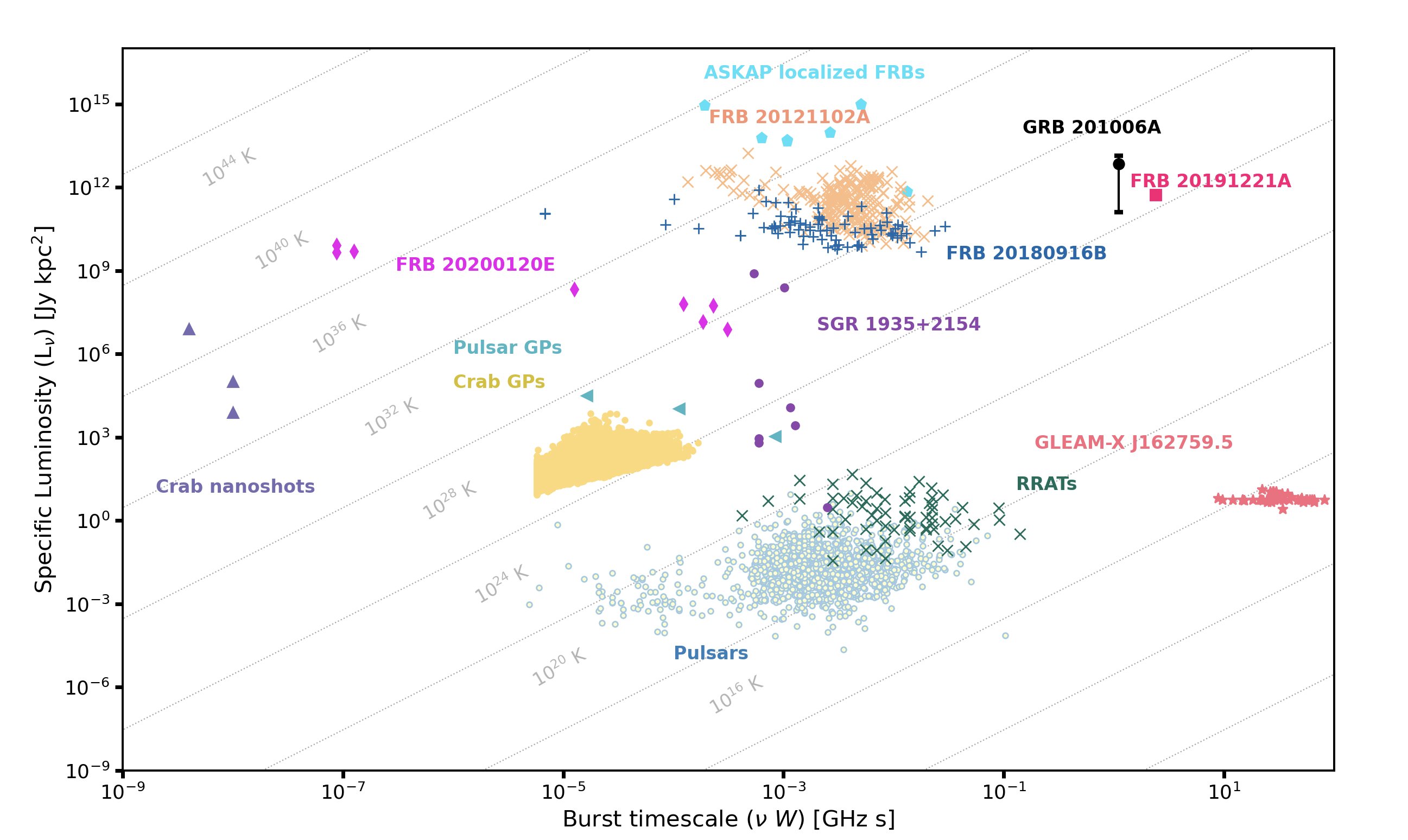}}
\caption{This figure illustrates the known populations of coherent radio transient sources with their specific luminosity on the y-axis and their characteristic burst timescale on the x-axis \citep{pietka2015, nimmo2022}. The radio transient associated with GRB 201006A is marked by the black cross assuming an intrinsic duration of 5 seconds.
\label{fig:tps}}
\end{figure*}

Assuming the radio transient has a DM of 770 pc cm$^{-3}$, there is an excess DM of 542 pc cm$^{-3}$ along this line of sight. Accounting for the DM range (740--800 pc cm$^{-3}$) and a 1$\sigma$ uncertainty of 0.02 from the observed scatter in the Macquart correlation, we use the Macquart correlation \citep{james2022} to calculate a redshift of $z = 0.58\pm0.06$, which is consistent with the redshift distribution of short GRBs \citep{DAvanzo14,rowlinson2013}. This redshift corresponds to a luminosity distance of 3.38 Gpc. Using the peak flux density from the 5 second dedispersed image at a DM of 770 pc cm$^{-3}$, $49\pm27$ mJy (corresponding to a fluence of $245\pm135$ Jy ms), and the redshift of $0.58\pm0.06$ we calculate a luminosity of $6.7^{+6.6}_{-4.4} \times 10^{32}$ erg s$^{-1}$ Hz$^{-1}$. We can then check where the source lies in the transient phase space. We estimate the burst timescale to 0.72~GHz~s and Figure \ref{fig:tps} shows the coherent transients \citep{pietka2015, nimmo2022} along with GRB~201006A. This radio source is comparable in luminosity to the population of well-localised FRBs \citep{prochaska2019}, but with a longer duration. We note the duration similarity between this radio flash and the $\sim$3 second FRB (20191221A), with a periodicity of 217 ms, detected by CHIME/FRB \citep{chime2022}. Several progenitor theories for FRBs invoke magnetars or energetic neutron stars (though these are typically expected to be spinning much slower than millisecond magnetars) \citep{petroff2022}, which could also be consistent with the newborn millisecond magnetar in this scenario.

\section{Modelling the emission}
\label{sec:models}

In this section, we assume that the coherent radio flash is associated with GRB 201006A. We first consider the likelihood of the coherent radio emission escaping from the region surrounding the GRB. Then we consider different theoretical models for coherent radio emission from neutron star binary mergers.

\subsection{Propagation considerations for coherent radio emission}

The propagation of low frequency coherent radio emission is very dependent upon the medium that the emission is traveling through. Dense plasma surrounding the progenitor system may be opaque to low frequency radio emission, thus preventing detectable coherent radio emission from escaping the source. \cite{zhang2014} demonstrated that this emission can escape along the relativistic jet axis for compact binary mergers \citep[for further discussion regarding the surrounding plasma, see][]{bhardwaj2023}. Further propagation effects are considered in \cite{rowlinson2019b}. 

Here, we also consider the ejecta in the relativistic jet, which may further prevent emission from the central engine from escaping the source. GRBs are expected to have an afterglow from the reverse and forward shocks that the ejecta cause when ploughing into the ambient medium. The detection of low-frequency radio emission from the magnetar requires the shocked medium to be transparent to this relatively low frequency radiation, and our attribution of a component of the X-ray afterglow to the magnetar requires the normal afterglow emission to be subdominant. These two conditions impose significant constraints on the explosion energy and the ambient density of the GRB. Specifically, we assume the nominal redshift we derived and assign fairly standard shock parameters (1\% of the energy into magnetic fields and 10\% into ultra-relativistic electrons) and test both the rising and deceleration phase of the afterglow \citep{gao2013}. We then find that relatively standard isotropic energies of $10^{50}$--$10^{52}$ erg are allowed, provided that the ambient density is rather low, $n = 10^{-4}$--$10^{-3}$ cm$^{-3}$, and the initial Lorentz factor is not above 100. The density constraint is below that which we find on average for GRBs with well-measured afterglows \citep{aksulu2022}, but agrees well with the fact that this GRB is short and hostless.

\subsection{Constraining an X-ray flare from a black hole}

Models predicting coherent radio emission from a black hole are caused by a magnetised wind launched by an accretion event \citep{usov2000}. However, the accretion disks following binary neutron star mergers are expected to fall onto the black hole within the first few seconds \citep{rezzolla2011}. There is a chance of material being flung out on a highly elliptical orbit that can accrete onto the black hole at late times, powering an X-ray flare \citep{rosswog2007}. 

Using this black hole model \citep{usov2000,starling2020}, we can work backwards to predict the X-ray flare flux given an observed radio pulse. Given a radio peak flux density $49\pm27$ mJy, duration $5$ seconds, redshift $0.58 \pm 0.06 $ and an efficiency, for converting the released energy into radio emission, of $10^{-3}$ as assumed by \cite{usov2000}, we find the peak X-ray flux required to generate the radio pulse to be $(9.9 \pm 5.5) \times 10^{-11}$ erg cm$^{-2}$ s$^{-1}$. This is roughly two orders of magnitude greater than the observed flux in the {\it Swift} XRT light curve at the expected time of this radio flare after correcting for the dispersion delay ($\sim 10^{-12}$ erg cm$^{-2}$ s$^{-1}$). The predicted luminosity of the X-ray flare is shown as a magenta circle in Figure \ref{fig:restframe_lc}. For the X-ray flare to be detectable, the efficiency would need to be relatively high ($\sim 10^{-2}$). Hence, the magnetised wind launched from a black hole model prediction is not consistent with the observations. Therefore, we conclude that the radio flare from GRB 201006A is unlikely to originate from accretion onto a black hole.

\subsection{Modelling X-ray emission with the magnetar model}

\begin{figure}
\centerline{\includegraphics[width=0.5\textwidth]{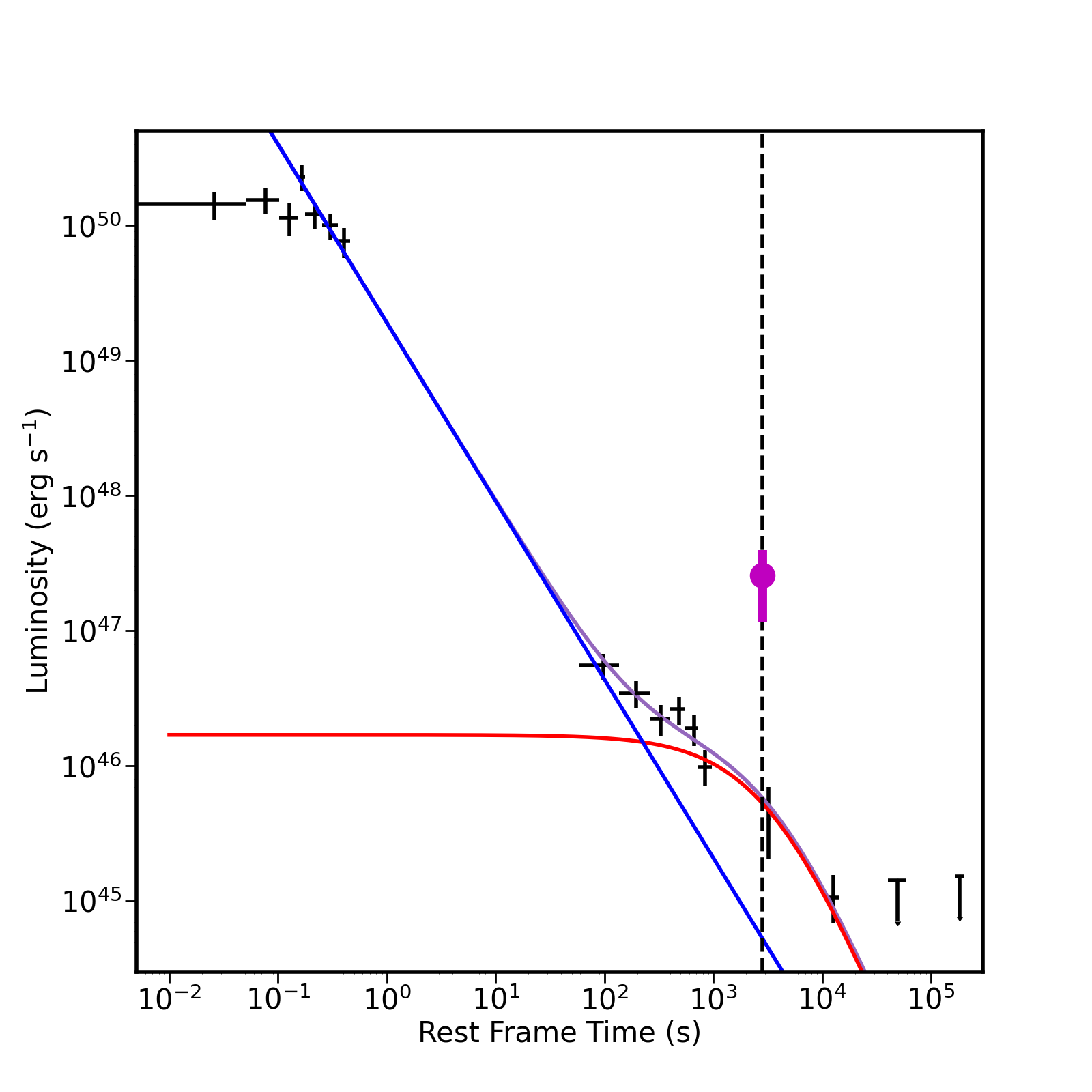}}
\caption{The black data points show the {\it Swift} BAT and XRT data, 0.2 -- 10 keV, for GRB 201006A at a redshift of 0.58 (obtained from the DM analysis). The red line shows the magnetar model component, the blue line shows the afterglow emission from the prompt pulse (likely due to curvature radiation) and the purple line shows the combined fit. The black dashed line shows the restframe time of emission of the observed coherent radio burst.
The magenta circle shows the predicted luminosity, and 1$\sigma$ errors, of an X-ray flare associated with the radio flash \citep{usov2000}. \label{fig:restframe_lc}}
\end{figure}

Instead of a black hole, we consider a millisecond magnetar origin for the observed emission, similar to the theorised progenitors of FRBs \citep{petroff2022}. Due to event rates, only a very small fraction of FRBs can come from millisecond magnetars formed via binary mergers, however more millisecond magnetars are also expected to be formed via superluminous supernovae and long GRBs \citep{kasen2010,metzger2011}. In previous works \citep[e.g. ][]{rowlinson2013}, it has been shown that the X-ray light curves of many short GRBs have a prolonged energy injection phase, which is consistent with spin-down emission from a newly formed magnetar.

\begin{figure}
\centerline{\includegraphics[width=0.44\textwidth]{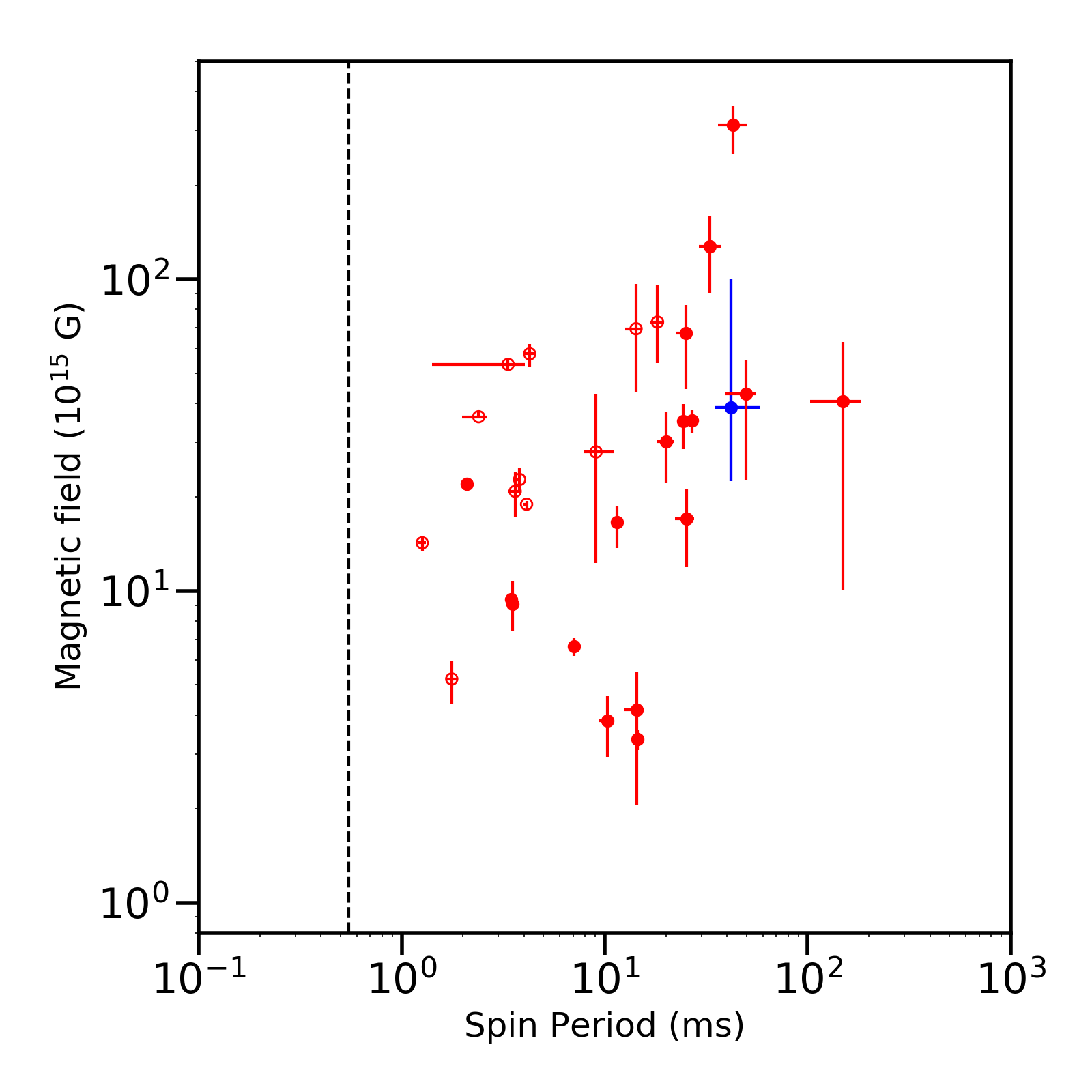}}
\caption{The red data points show the magnetic fields and initial spin periods of the population of magnetars fitted to the restframe X-ray light curves of short GRBs using the identical method to that presented in this paper \citep{rowlinson2013}. The blue data point shows the magnetar parameters fitted for GRB 201006A assuming a mass of 2.1 $M_{\odot}$ and the redshift of 0.58 obtained from the DM analysis. All data points are assuming a factor of $f = 3.45$ to account for beaming and efficiency \citep{rowlinson2019}. The dashed black line shows the spin break-up for a 2.1 $M_{\odot}$ neutron star.
\label{fig:BP_plot}}
\end{figure}

The 0.3--10 keV BAT-XRT light curve of GRB 201006A was transformed into a 1--10,000 keV restframe light curve, using the redshift of 0.58 obtained from the DM analysis, using existing methods \citep{rowlinson2013}, and is shown in Figure \ref{fig:restframe_lc}. Using the magnetar spin-down relations \citep{zhang2001} and assuming a factor, $f\sim3.45$, which encompasses the beaming angle and efficiency uncertainties \citep{rowlinson2014}, and assuming a remnant mass of 2.1$M_{\odot}$, we fit the restframe light curve of GRB 201006A using existing methods \citep{rowlinson2013, rowlinson2014}. We find that the restframe X-ray light curve of GRB 201006A is well fit by a spinning down magnetar, with a magnetic field $39^{+61}_{-16} \times 10^{15}$ G and initial spin period of $42^{+17}_{-7}$ ms. In Figure \ref{fig:BP_plot}, we show this magnetar in relation to other short GRB magnetar candidates fitted using the same method \citep{rowlinson2013}. The magnetar fitted using the X-ray data from GRB 201006A is consistent with the rest of the population of magnetars fitted using X-ray data from short GRBs. 

We can predict the spin period of the newly formed magnetar at the rest frame time the radio flare was emitted (48 minutes after formation) using the fitted initial spin period and magnetic field strength \citep{rowlinson2017}. We find the spin period of the magnetar at this time to be $481^{+1236}_{-280}$ ms. Interestingly, this is comparable to the 217 ms intrinsic periodicity observed during the 3 second FRB \citep{chime2022}.

\subsection{Coherent radio emission expectations}

\subsubsection{Emission from a stable magnetar}
A newborn magnetar may be emitting in a similar manner to standard pulsars, such that a proportion of its spin-down energy is converted to radio emission in addition to the X-ray emission observed. Assuming a standard neutron star efficiency \citep{taylor1993} ($\epsilon_{r}= 10^{-4}$), the predicted radio flux density, $f_{\nu}$ is given by \citep{totani2013}:
\begin{eqnarray}
f_{\nu} = \frac{1}{\nu_{\rm obs}}\frac{\epsilon_{r} L}{4 \pi D^2 } = 3.2 \times 10^{10} \frac{\epsilon_{\rm r}}{\nu_{\rm MHz} D_{\rm Gpc}^2}   B^2_{15} R^6_6 P^{-4}_{-3} ~{\rm Jy} 
\label{eqn:totaniFlx}
\end{eqnarray}
where $\nu_{\rm obs}$ is the observing frequency in Hz, $\nu_{\rm MHz}$ is the observing frequency in MHz, $D$ is the distance in cm, $D_{Gpc}$ is the distance in Gpc, $B_{15}$ is the magnetic field of the magnetar in units of $10^{15}$ G, $P_{-3}$ is the spin period of the magnetar in ms, and $R_{6}$ is the radius of the magnetar in $10^6$ cm. By combining this with the magnetar central engine model developed by \cite{zhang2001}, we can derive the time evolution of the predicted radio emission. 
\begin{eqnarray}
f_{\nu}(t) = 3.2 \times 10^{10} \frac{\epsilon_{\rm r}}{\nu_{\rm MHz} D_{\rm Gpc}^2} B^2_{15} R^6_6 P^{-4}_{-3} ~{\rm Jy} \nonumber \\
= 3.2 \times 10^{10} \frac{\epsilon_{\rm r}}{\nu_{\rm MHz} D_{\rm Gpc}^2} L_{49}(t) ~ {\rm Jy}
\label{eqn:totaniFlx2} \\
L_{49}(t) = L_{0,49} \left( 1 + \frac{t}{T}   \right)^{-2} \label{eqn:totaniFlx3}
\end{eqnarray}
where $L_{49}$ is the bolometric luminosity in $10^{49}$ erg s$^{-1}$, $L_{0,49}$ is the initial luminosity,  and $T$ is the duration of the plateau phase in seconds. The initial luminosity and plateau duration are given by (including the beaming and efficiency factor, $f$)
\begin{eqnarray}
    L_{0,49} = B^2_{15} R^6_6 P^{-4}_{0,-3} f \label{eqn:zhangL} \\
    T = 2.05 \times 10^3 M_{1.4} B_{15}^{-2} P_{0,-3}^{2} R_6^{-4} \label{eqn:zhangT}
\end{eqnarray}
where $M_{1.4} = 1.4 M_{\odot}$ and $P_{0,-3}$ is the initial spin period of the magnetar. The radius of the magnetar is expected to settle to approximately 10 km ($R_6$=1) within the first few seconds \citep{metzger2011}. Inputting the fitted magnetar parameters and the emission time of 48 minutes after formation of the magnetar, given a redshift of 0.58 (corresponding to a luminosity distance of 3.38 Gpc) and the observing frequency $\nu_{\rm MHz} = 144$ MHz, we find the predicted flux density is $627^{+247}_{-429}$ mJy. This is inconsistent with the non-detection in the deep image of the field of GRB 201006A with a 3$\sigma$ upper limit of 3 mJy.  A rapidly rotating neutron star may also give off giant pulses \citep{karuppusamy2010} that, with increased beaming or efficiency, could be a progenitor of FRBs. We conclude that the observed $49\pm27$ mJy radio flare could be consistent with a single pulse from the magnetar with a lower efficiency of $\epsilon_{\rm r} = 10^{-5}$.

The discovery of bright radio bursts from galactic magnetar SGR 1935+2154 \citep{bochenek2020,chime2020,kirsten2021} shows that magnetars produce magnetically powered coherent radio emission similar to FRBs. The magnetar in our scenario could also power an FRB via release of a proportion of its magnetic energy as a giant flare \citep{lyubarsky2014,beloborodov2017}. Typically, the emission mechanism proposed is that the giant flare shocks the surrounding plasma and produces radio emission \citep{lyubarsky2014}. Taking equation 14 from \cite{gourdji2020} for the predicted isotropic energy, $E_{\rm pred}$, and the observed isotropic energy from the radio emission, $E_{\rm obs}$, we can calculate the fraction of the magnetic field of the magnetar contained within in the flare, $b$, using
\begin{eqnarray}
b^2 = \frac{E_{\rm obs}}{E_{\rm pred}},
\label{eqn:lyubarsky} \\
E_{\rm pred} = \frac{\eta B^2 R^2 n m_e c^3 b^2\Delta t}{16 p \xi}, \label{eqn:lyubarskyEnergy} \\
E_{\rm obs} = 4 \pi \mathcal{F} D^2 \nu, \label{eqn:obsEnergy}
\end{eqnarray}
where $\eta$ is the fraction of energy that escapes the plasma, $B$ is the magnetic field in G, $R$ is the radius of the neutron star in cm, $n$ is the particle density in the shock in g cm$^{-3}$, $m_e$ is the mass of the electron in cgs units, $c$ is the speed of light in cgs units, $\Delta t$ is the duration of the flare, $p$ is the pressure in the nebula surrounding the neutron star and $\xi$ is the fraction of particles that lose their energy before entering the nebula. In this analysis, we assume $\eta = \xi$ \citep{gourdji2020}, $p=10^{-8}$ cm$^{-3}$ and $n=4\times10^{-24}$ g cm$^{-3}$ \citep{lyubarsky2014}. Inputting the parameters for the observed radio flare and the fitted magnetar, we find $b = 4.6^{+6.6}_{-2.2}$. Therefore, the observed energy is significantly larger than the predicted energy for the model. We note that this model assumes that there is a nebula surrounding the magnetar and it is unclear if this would be the case for a newly formed magnetar. If there is only a negligible nebula, then $p \rightarrow 0$ and the predicted energy would rapidly increase. Therefore, given standard assumptions, it is unlikely that the nebula shock emission model can explain the observed radio source. We note that this magnetar has a very large magnetic energy reservoir that could produce FRBs via a different emission mechanism or be in a significantly lower density environment.

\subsubsection{Emission from an unstable magnetar}
Depending upon the mass of the newly formed magnetar and the equation of state of nuclear matter, the magnetar will either be stable or unstable. If the young magnetar is unstable it will collapse to form a black hole at later times and this is dependent upon the equation of state of nuclear matter. Initially the supramassive neutron star is typically supported via its rapid rotation \citep{ozel2010} but, as it quickly loses rotational energy, it can reach a critical point where it can no longer support its own mass, collapsing to form a black hole. This collapse is typically expected within the first few hours of the formation of the millisecond magnetar and the collapse time can be used to constrain the nuclear equation of state \citep{lasky2014,beniamini2021}. This scenario has been used to tentatively suggest FRB 20190425A resulted from the collapse of a magnetar created during GW190425 at 2.5 hr following the binary neutron star merger \citep{moroianu2022}. The magnetar fitted to GRB 201006A has a relatively slow spin period and is therefore unlikely to be supported via its rotation. However, it has a very large magnetic field ($4 \times 10^{16}$ G), thus it could be a supramassive neutron star supported via its magnetic fields with an expected collapse time of $\sim$years \citep{suvorov2022}. However, it has also been shown that for magnetars, with magnetic fields up to $3\times 10^{16}$ G, all neutron star remnants are expected to collapse within $4.4 \times 10^{4}$ seconds \citep{ravi2014}. Thus, the candidate radio flash detected at $\sim$4600 seconds after formation would be a very reasonable time for a magnetar collapse.

The magnetar model fitted to GRB 201006A assumes a stable magnetar is formed as there is no steep decay phase following the plateau \citep{rowlinson2013}. A different interpretation suggests that the shallow decay following the plateau phase is caused by afterglow emission and the steep decay following the plateau (signifying a collapsing magnetar) is `hidden' under this afterglow component \citep{lu2015}. Thus, we do not rule out the presence of an unstable magnetar that collapses to form a black hole at the end of the plateau phase.

When a magnetar collapses to form a black hole, there will be a massive reconnection event when the magnetic field collapses and this is predicted to produce a copious amount of coherent radio emission \citep{falcke2014,zhang2014}. The predicted flux density of this emission, assuming it is isotropic, is given by \citep{zhang2014,rowlinson2019}:
\begin{eqnarray}
    f_{\nu} = -\frac{10^{-23} \epsilon E_B}{ 4 \pi D^2 \tau} (\alpha +1) \nu_p^{-(\alpha +1)} ~ \frac{\nu_{obs}^{\alpha}}{(1+z)} \frac{\tau}{t_{\rm int}} ~{\rm Jy}, 
\end{eqnarray}
where $\epsilon$ is the fraction of the energy that is expected to be converted into coherent radio emission (assumed to be $\epsilon = 10^{-6}$), $E_{B}$ is the amount of energy available in the magnetic field of the magnetar (given by $E_{B} = 1.7 \times 10^{47} B^{2}_{15} R^{3}_{6}$ erg), $\tau$ is the intrinsic width of the emission in seconds (taken to be 5 seconds as the maximum duration of this event determined via the DM analysis) \citep{falcke2014}, $\alpha$ is the spectral index of the radio emission (assumed to be -2), $\nu_{p}$ is the plasma frequency given by $\nu_{p} = 3.8\times10^{6} \frac{L_{\gamma}^{0.5}}{0.01\gamma R_{6}}$ ($L_{\gamma}$ is the luminosity of the gamma-ray burst and $\gamma$ is the Lorentz factor of the jet) and $t_{int}$ is the integration time of the image in seconds (also taken to be 5 seconds). The fluence of GRB 201006A is $(3.33 \pm 0.38)\times 10^{-7}$ erg cm$^{-2}$ (10--1000 keV) as determined by {\it Fermi}, corresponding to a flux of $f = \frac{fluence}{t_{90}} = \frac{3.33\times 10^{-7}}{1.7} = 1.96\times 10^{-7}$ erg cm$^{-2}$ s$^{-1}$ \citep{hamburg2020}. The 1-10000 keV luminosity of GRB 201006A is $5.1\times 10^{50}$ erg s$^{-1}$ at a redshift of 0.58. The luminosity is then extrapolated to scale the flux density prediction to different redshifts. For $\gamma$, we take a representative Lorentz factor of relativistic jets in short GRBs as $\gamma =1000$ \citep{ackermann2010}.

\begin{figure}
\centerline{\includegraphics[width=0.5\textwidth]{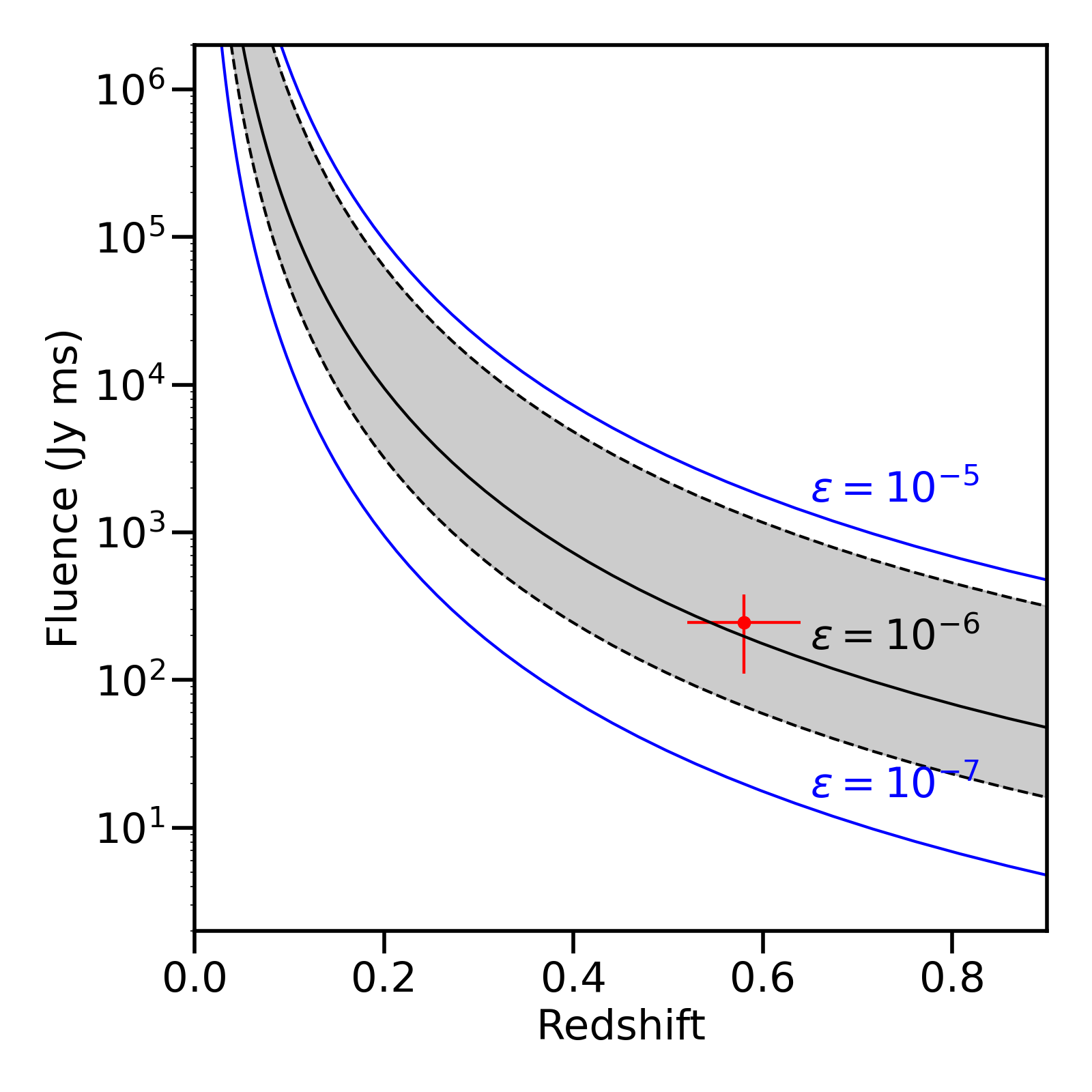}}
\caption{The solid black line shows the predicted radio fluence at 144 MHz for a collapsing magnetar \citep{zhang2014,rowlinson2019} using the fitted magnetar parameters for GRB 201006A and assuming an efficiency of $10^{-6}$ in the conversion of the released energy into coherent radio emission. The shaded grey region shows the 1$\sigma$ uncertainty in this prediction given by the fitted uncertainties on the magnetar parameters. The blue solid lines show the predictions assuming efficiencies of $10^{-5}$ and $10^{-7}$. The red data point shows the fluence of the observed radio emission at the inferred redshift of $0.58\pm0.06$.
\label{fig:magnetar_radio}}
\end{figure}

For a redshift of 0.58, the predicted fluence of radio emission from the modelled magnetar collapsing to a black hole is $208^{+1380}_{-70} $ Jy ms. Given an observed flux density of $49\pm27$ mJy in a 5 second image, the observed fluence is $245\pm135$ Jy ms. A number of assumptions are made in this model, particularly for the efficiency $\epsilon$ and the intrinsic duration of the radio flare, and the quoted uncertainties are for 1$\sigma$. The prediction as a function of redshift is plotted in Figure \ref{fig:magnetar_radio}.  We compare the predictions to the observed fluence of the candidate radio flash following GRB 201006A and conclude that the radio emission would be consistent with the millisecond magnetar collapsing to form a black hole 48 minutes (in the rest frame) following its formation. 

\section{Conclusions}

We conducted a 2 hour rapid response LOFAR observation of the short GRB 201006A. Following time slicing and a transient search, we detected a flash of coherent radio emission 27 arcsec from the GRB location and 76.5 minutes after the GRB. While this flash was 3.8$\sigma$ offset from the GRB location when considering statistical position uncertainties, we argue that it is likely to be associated with GRB 201006A as the probability of being an unrelated transient is $10^{-6}$. If associated with GRB 201006A, we expect this radio flash to be highly dispersed and we conducted an image plane dedispersion search on the radio data. We attained a tentative detection of dispersion, 2.4$\sigma$, which corresponds to a redshift of $z=0.58\pm0.06$. 

Assuming the radio flash is associated with GRB 201006A, we considered the origin of the coherent radio emission. If the central engine were a black hole, the emission would be likely associated with an accretion event that should have been detectable in the X-ray light curve. As there is no X-ray flare in the light curve at the time of the radio flare, we conclude a black hole remnant is unlikely. If instead the engine were a millisecond magnetar, we would expect coherent radio emission at this time. We fitted the X-ray light curve to determine the magnetar properties and used them to predict the coherent radio emission from different emission models. We find that the radio flare would be consistent with being a single pulse from the newborn magnetar or with the final collapse of the magnetar to a black hole.

If associated, this radio flash would be the first detection of coherent radio emission from a GRB. However, due to the positional offset, we cannot confidently confirm this association. We have demonstrated in this paper the exciting physics we can probe with coherent radio emission from these sources and further rapid response observations of GRBs with radio telescopes are therefore very strongly encouraged. Compact binary mergers are also predicted to produce copious gravitational waves that can be detected in the nearby Universe but these sources are typically very poorly localised. Identifying coherent radio emission by obtaining rapid observations of a gravitational wave detection \citep{gourdji2023} would enable us to localise the event to a few arcseconds as opposed to the thousands of degrees expected from the gravitational wave signal \citep{chu2016}. Additionally, this enables physical constraints on the key physics behind neutron star mergers, such as probing the nature of the merger remnant. The confirmation of magnetar remnants formed from neutron star mergers, especially combined with observed masses from gravitational wave signals \citep{lasky2014,beniamini2021}, will have significant consequences regarding the highly debated equation of state of nuclear matter by providing evidence of supramassive neutron stars bridging the mass gap to the lightest black holes.

\section*{Acknowledgements}

We thank the LOFAR Radio Observatory for implementing the rapid response mode and for supporting our observations. We thank J. Hessels, M. Patel and P. Evans for useful discussions. 

AR acknowledges funding from the NWO Aspasia grant (number: 015.016.033). IdR \& AR acknowledge funding from the NWA CORTEX grant (NWA.1160.18.316) of the research programme NWA-ORC which is (partly) financed by the Dutch Research Council (NWO). RLCS acknowledges support from STFC, the ASTRON/JIVE Helena Kluyver visitor program and a Leverhulme Research Project award. AH is supported by an STFC Studentship.
GEA acknowledges the support of an Australian Research Council Discovery Early Career Researcher Award (project number DE180100346).
KW acknowledges support through a UK Research and Innovation Future Leaders Fellowship (MR/T044136/1; PI Dr.~B. Simmons). KG acknowledges support through Australian Research Council Discovery Project DP200102243. 

This paper is based on data obtained with the International LOFAR Telescope (ILT) under project code LC14\_004 (PI: Starling). LOFAR \citep{vanhaarlem2013} is the Low Frequency Array designed and constructed by ASTRON. It has observing, data processing, and data storage facilities in several countries, that are owned by various parties (each with their own funding sources), and that are collectively operated by the ILT foundation under a joint scientific policy. The ILT resources have benefited from the following recent major funding sources: CNRS-INSU, Observatoire de Paris and Universit{\'e} d'Orl{\`e}ans, France; BMBF, MIWF-NRW, MPG, Germany; Science Foundation Ireland (SFI), Department of Business, Enterprise and Innovation (DBEI), Ireland; NWO, The Netherlands; The Science and Technology Facilities Council, UK; Ministry of Science and Higher Education, Poland.

This work made use of data supplied by the UK {\it Swift} Science Data Centre at the University of Leicester and the {\it Swift} satellite. {\it Swift}, launched in November 2004, is a NASA mission in partnership with the Italian Space Agency and the UK Space Agency. {\it Swift} is managed by NASA Goddard. Penn State University controls science and flight operations from the Mission Operations Center in University Park, Pennsylvania. Los Alamos National Laboratory provides gamma-ray imaging analysis.

\section*{Data Availability}

The LOFAR data used are available in the LOFAR Long Term Archive at \url{https://lta.lofar.eu} under the Cycle 14 project title LC14{\_}004 with SAS Id numbers 796450 \& 797382 (target data) and 796452 \& 797386 (calibrator data).
{\it Swift} data are available through the UK Swift Science Data Centre \url{www.swift.ac.uk}.

The scripts underlying this article are available in Zenodo, at \url{https://zenodo.org/record/13833551}. 

The LOFAR calibration pipeline, {\sc LINC} is available at \url{https://linc.readthedocs.io/en/latest/index.htm}. {\sc WSClean}, used to create the images used in this paper, is available at \url{https://wsclean.readthedocs.io/en/latest/}.
The LOFAR Transients Pipeline ({\sc TraP}) is available here \url{https://docs.transientskp.org}. The Live Pulse Finder ({\sc LPF}) used to find sources in the dedispersed images is available here \url{https://github.com/transientskp/lpf}. {\sc LORDS}, used to conduct the dedispersion imaging, will be available at \url{https://git.astron.nl/rajwade/lordss} when it is made publicly available at a later date \citep{rajwade}.



\bibliographystyle{mnras}
\bibliography{references} 

\begin{thebibliography}{}
\makeatletter
\relax
\def\mn@urlcharsother{\let\do\@makeother \do\$\do\&\do\#\do\^\do\_\do\%\do\~}
\def\mn@doi{\begingroup\mn@urlcharsother \@ifnextchar [ {\mn@doi@} {\mn@doi@[]}}
\def\mn@doi@[#1]#2{\def\@tempa{#1}\ifx\@tempa\@empty \href {http://dx.doi.org/#2} {doi:#2}\else \href {http://dx.doi.org/#2} {#1}\fi \endgroup}
\def\mn@eprint#1#2{\mn@eprint@#1:#2::\@nil}
\def\mn@eprint@arXiv#1{\href {http://arxiv.org/abs/#1} {{\tt arXiv:#1}}}
\def\mn@eprint@dblp#1{\href {http://dblp.uni-trier.de/rec/bibtex/#1.xml} {dblp:#1}}
\def\mn@eprint@#1:#2:#3:#4\@nil{\def\@tempa {#1}\def\@tempb {#2}\def\@tempc {#3}\ifx \@tempc \@empty \let \@tempc \@tempb \let \@tempb \@tempa \fi \ifx \@tempb \@empty \def\@tempb {arXiv}\fi \@ifundefined {mn@eprint@\@tempb}{\@tempb:\@tempc}{\expandafter \expandafter \csname mn@eprint@\@tempb\endcsname \expandafter{\@tempc}}}

\bibitem[\protect\citeauthoryear{{Abbott} et~al.,}{{Abbott} et~al.}{2017a}]{abbott2017}
{Abbott} B.~P.,  et~al., 2017a, \mn@doi [Phys. Rev. Lett.] {10.1103/PhysRevLett.119.161101}, \href {https://ui.adsabs.harvard.edu/abs/2017PhRvL.119p1101A} {119, 161101}

\bibitem[\protect\citeauthoryear{{Abbott} et~al.,}{{Abbott} et~al.}{2017b}]{abbott2017b}
{Abbott} B.~P.,  et~al., 2017b, \mn@doi [\apjl] {10.3847/2041-8213/aa920c}, \href {https://ui.adsabs.harvard.edu/abs/2017ApJ...848L..13A} {848, L13}

\bibitem[\protect\citeauthoryear{{Ackermann} et~al.,}{{Ackermann} et~al.}{2010}]{ackermann2010}
{Ackermann} M.,  et~al., 2010, \mn@doi [\apj] {10.1088/0004-637X/716/2/1178}, \href {https://ui.adsabs.harvard.edu/abs/2010ApJ...716.1178A} {716, 1178}

\bibitem[\protect\citeauthoryear{{Aksulu}, {Wijers}, {van Eerten}  \& {van der Horst}}{{Aksulu} et~al.}{2022}]{aksulu2022}
{Aksulu} M.~D.,  {Wijers} R.~A.~M.~J.,  {van Eerten} H.~J.,   {van der Horst} A.~J.,  2022, \mn@doi [\mnras] {10.1093/mnras/stac246}, \href {https://ui.adsabs.harvard.edu/abs/2022MNRAS.511.2848A} {511, 2848}

\bibitem[\protect\citeauthoryear{{Amati}, {Guidorzi}, {Frontera}, {Della Valle}, {Finelli}, {Landi}  \& {Montanari}}{{Amati} et~al.}{2008}]{Amati08}
{Amati} L.,  {Guidorzi} C.,  {Frontera} F.,  {Della Valle} M.,  {Finelli} F.,  {Landi} R.,   {Montanari} E.,  2008, \mn@doi [\mnras] {10.1111/j.1365-2966.2008.13943.x}, \href {https://ui.adsabs.harvard.edu/abs/2008MNRAS.391..577A} {391, 577}

\bibitem[\protect\citeauthoryear{{Anderson} et~al.,}{{Anderson} et~al.}{2021}]{anderson2021}
{Anderson} G.~E.,  et~al., 2021, \mn@doi [\pasa] {10.1017/pasa.2021.15}, \href {https://ui.adsabs.harvard.edu/abs/2021PASA...38...26A} {38, e026}

\bibitem[\protect\citeauthoryear{{Balsano} et~al.,}{{Balsano} et~al.}{1998}]{balsano1998}
{Balsano} R.~J.,  et~al., 1998, in {Meegan} C.~A.,  {Preece} R.~D.,   {Koshut} T.~M.,  eds,  American Institute of Physics Conference Series Vol. 428, Gamma-Ray Bursts, 4th Hunstville Symposium. pp 585--589, \mn@doi{10.1063/1.55382}

\bibitem[\protect\citeauthoryear{{Bannister}, {Murphy}, {Gaensler}  \& {Reynolds}}{{Bannister} et~al.}{2012}]{bannister2012}
{Bannister} K.~W.,  {Murphy} T.,  {Gaensler} B.~M.,   {Reynolds} J.~E.,  2012, \mn@doi [\apj] {10.1088/0004-637X/757/1/38}, \href {https://ui.adsabs.harvard.edu/abs/2012ApJ...757...38B} {757, 38}

\bibitem[\protect\citeauthoryear{{Barthelmy} et~al.,}{{Barthelmy} et~al.}{2005}]{barthelmy2005}
{Barthelmy} S.~D.,  et~al., 2005, \mn@doi [\ssr] {10.1007/s11214-005-5096-3}, \href {http://adsabs.harvard.edu/abs/2005SSRv..120..143B} {120, 143}

\bibitem[\protect\citeauthoryear{{Beloborodov}}{{Beloborodov}}{2017}]{beloborodov2017}
{Beloborodov} A.~M.,  2017, \mn@doi [\apjl] {10.3847/2041-8213/aa78f3}, \href {https://ui.adsabs.harvard.edu/abs/2017ApJ...843L..26B} {843, L26}

\bibitem[\protect\citeauthoryear{{Beniamini} \& {Lu}}{{Beniamini} \& {Lu}}{2021}]{beniamini2021}
{Beniamini} P.,  {Lu} W.,  2021, \mn@doi [\apj] {10.3847/1538-4357/ac1678}, \href {https://ui.adsabs.harvard.edu/abs/2021ApJ...920..109B} {920, 109}

\bibitem[\protect\citeauthoryear{{Bhardwaj}, {Palmese}, {Maga{\~n}a Hernandez}, {D'Emilio}  \& {Morisaki}}{{Bhardwaj} et~al.}{2023}]{bhardwaj2023}
{Bhardwaj} M.,  {Palmese} A.,  {Maga{\~n}a Hernandez} I.,  {D'Emilio} V.,   {Morisaki} S.,  2023, \mn@doi [arXiv e-prints] {10.48550/arXiv.2306.00948}, \href {https://ui.adsabs.harvard.edu/abs/2023arXiv230600948B} {p. arXiv:2306.00948}

\bibitem[\protect\citeauthoryear{{Bochenek}, {Ravi}, {Belov}, {Hallinan}, {Kocz}, {Kulkarni}  \& {McKenna}}{{Bochenek} et~al.}{2020}]{bochenek2020}
{Bochenek} C.~D.,  {Ravi} V.,  {Belov} K.~V.,  {Hallinan} G.,  {Kocz} J.,  {Kulkarni} S.~R.,   {McKenna} D.~L.,  2020, \mn@doi [\nat] {10.1038/s41586-020-2872-x}, \href {https://ui.adsabs.harvard.edu/abs/2020Natur.587...59B} {587, 59}

\bibitem[\protect\citeauthoryear{{Burrows} et~al.,}{{Burrows} et~al.}{2005}]{burrows2005}
{Burrows} D.~N.,  et~al., 2005, \mn@doi [\ssr] {10.1007/s11214-005-5097-2}, \href {https://ui.adsabs.harvard.edu/abs/2005SSRv..120..165B} {120, 165}

\bibitem[\protect\citeauthoryear{{CHIME/FRB Collaboration} et~al.,}{{CHIME/FRB Collaboration} et~al.}{2020}]{chime2020}
{CHIME/FRB Collaboration} et~al., 2020, \mn@doi [\nat] {10.1038/s41586-020-2863-y}, \href {https://ui.adsabs.harvard.edu/abs/2020Natur.587...54C} {587, 54}

\bibitem[\protect\citeauthoryear{{Carbone} et~al.,}{{Carbone} et~al.}{2018}]{carbone2018}
{Carbone} D.,  et~al., 2018, \mn@doi [Astronomy and Computing] {10.1016/j.ascom.2018.02.003}, \href {https://ui.adsabs.harvard.edu/abs/2018A&C....23...92C} {23, 92}

\bibitem[\protect\citeauthoryear{{Chime/Frb Collaboration} Andersen et~al.,}{{Chime/Frb Collaboration} et~al.}{2022}]{chime2022}
{Chime/Frb Collaboration} Andersen B.~C.,  et~al., 2022, \mn@doi [\nat] {10.1038/s41586-022-04841-8}, \href {https://ui.adsabs.harvard.edu/abs/2022Natur.607..256C} {607, 256}

\bibitem[\protect\citeauthoryear{{Chu}, {Howell}, {Rowlinson}, {Gao}, {Zhang}, {Tingay}, {Bo{\"e}r}  \& {Wen}}{{Chu} et~al.}{2016}]{chu2016}
{Chu} Q.,  {Howell} E.~J.,  {Rowlinson} A.,  {Gao} H.,  {Zhang} B.,  {Tingay} S.~J.,  {Bo{\"e}r} M.,   {Wen} L.,  2016, \mn@doi [\mnras] {10.1093/mnras/stw576}, \href {https://ui.adsabs.harvard.edu/abs/2016MNRAS.459..121C} {459, 121}

\bibitem[\protect\citeauthoryear{{Condon}}{{Condon}}{1997}]{condon1997}
{Condon} J.~J.,  1997, \mn@doi [\pasp] {10.1086/133871}, \href {https://ui.adsabs.harvard.edu/abs/1997PASP..109..166C} {109, 166}

\bibitem[\protect\citeauthoryear{{Curtin} et~al.,}{{Curtin} et~al.}{2022}]{curtin2022}
{Curtin} A.~P.,  et~al., 2022, \mn@doi [arXiv e-prints] {10.48550/arXiv.2208.00803}, \href {https://ui.adsabs.harvard.edu/abs/2022arXiv220800803C} {p. arXiv:2208.00803}

\bibitem[\protect\citeauthoryear{{D'Avanzo} et~al.,}{{D'Avanzo} et~al.}{2014}]{DAvanzo14}
{D'Avanzo} P.,  et~al., 2014, \mn@doi [\mnras] {10.1093/mnras/stu994}, \href {https://ui.adsabs.harvard.edu/abs/2014MNRAS.442.2342D} {442, 2342}

\bibitem[\protect\citeauthoryear{{Demorest}, {Pennucci}, {Ransom}, {Roberts}  \& {Hessels}}{{Demorest} et~al.}{2010}]{demorest2010}
{Demorest} P.~B.,  {Pennucci} T.,  {Ransom} S.~M.,  {Roberts} M.~S.~E.,   {Hessels} J.~W.~T.,  2010, \mn@doi [\nat] {10.1038/nature09466}, \href {https://ui.adsabs.harvard.edu/abs/2010Natur.467.1081D} {467, 1081}

\bibitem[\protect\citeauthoryear{{Evans} et~al.,}{{Evans} et~al.}{2009}]{Evans09}
{Evans} P.~A.,  et~al., 2009, \mn@doi [\mnras] {10.1111/j.1365-2966.2009.14913.x}, \href {https://ui.adsabs.harvard.edu/abs/2009MNRAS.397.1177E} {397, 1177}

\bibitem[\protect\citeauthoryear{{Falcke} \& {Rezzolla}}{{Falcke} \& {Rezzolla}}{2014}]{falcke2014}
{Falcke} H.,  {Rezzolla} L.,  2014, \mn@doi [\aap] {10.1051/0004-6361/201321996}, \href {https://ui.adsabs.harvard.edu/abs/2014A&A...562A.137F} {562, A137}

\bibitem[\protect\citeauthoryear{{Fijma} et~al.,}{{Fijma} et~al.}{2023}]{fijma2023}
{Fijma} S.,  et~al., 2023, \mn@doi [arXiv e-prints] {10.48550/arXiv.2306.16383}, \href {https://ui.adsabs.harvard.edu/abs/2023arXiv230616383F} {p. arXiv:2306.16383}

\bibitem[\protect\citeauthoryear{{Fong} et~al.,}{{Fong} et~al.}{2022}]{Fong22}
{Fong} W.-f.,  et~al., 2022, \mn@doi [\apj] {10.3847/1538-4357/ac91d0}, \href {https://ui.adsabs.harvard.edu/abs/2022ApJ...940...56F} {940, 56}

\bibitem[\protect\citeauthoryear{{Gao}, {Lei}, {Zou}, {Wu}  \& {Zhang}}{{Gao} et~al.}{2013}]{gao2013}
{Gao} H.,  {Lei} W.-H.,  {Zou} Y.-C.,  {Wu} X.-F.,   {Zhang} B.,  2013, \mn@doi [\nar] {10.1016/j.newar.2013.10.001}, \href {https://ui.adsabs.harvard.edu/abs/2013NewAR..57..141G} {57, 141}

\bibitem[\protect\citeauthoryear{{Gehrels} et~al.,}{{Gehrels} et~al.}{2004}]{gehrels2004}
{Gehrels} N.,  et~al., 2004, \mn@doi [\apj] {10.1086/422091}, \href {https://ui.adsabs.harvard.edu/abs/2004ApJ...611.1005G} {611, 1005}

\bibitem[\protect\citeauthoryear{{Goad}, {Osborne}, {Beardmore}, {Evans}  \& {Swift-XRT Team.}}{{Goad} et~al.}{2020}]{goad2020}
{Goad} M.~R.,  {Osborne} J.~P.,  {Beardmore} A.~P.,  {Evans} P.~A.,   {Swift-XRT Team.} 2020, GRB Coordinates Network, \href {https://ui.adsabs.harvard.edu/abs/2020GCN.28562....1G} {28562, 1}

\bibitem[\protect\citeauthoryear{{Gourdji}, {Rowlinson}, {Wijers}  \& {Goldstein}}{{Gourdji} et~al.}{2020}]{gourdji2020}
{Gourdji} K.,  {Rowlinson} A.,  {Wijers} R.~A.~M.~J.,   {Goldstein} A.,  2020, \mn@doi [\mnras] {10.1093/mnras/staa2128}, \href {https://ui.adsabs.harvard.edu/abs/2020MNRAS.497.3131G} {497, 3131}

\bibitem[\protect\citeauthoryear{{Gourdji}, {Rowlinson}, {Wijers}, {Broderick}  \& {Shulevski}}{{Gourdji} et~al.}{2023}]{gourdji2023}
{Gourdji} K.,  {Rowlinson} A.,  {Wijers} R.,  {Broderick} J.,   {Shulevski} A.,  2023, \mn@doi [arXiv e-prints] {10.48550/arXiv.2303.11555}, \href {https://ui.adsabs.harvard.edu/abs/2023arXiv230311555G} {p. arXiv:2303.11555}

\bibitem[\protect\citeauthoryear{{Gropp} et~al.,}{{Gropp} et~al.}{2020}]{gropp2020}
{Gropp} J.~D.,  et~al., 2020, GRB Coordinates Network, \href {https://ui.adsabs.harvard.edu/abs/2020GCN.28560....1G} {28560, 1}

\bibitem[\protect\citeauthoryear{{Hamburg}, {Meegan}  \& {Fermi GBM Team}}{{Hamburg} et~al.}{2020}]{hamburg2020}
{Hamburg} R.,  {Meegan} C.,   {Fermi GBM Team} 2020, GRB Coordinates Network, \href {https://ui.adsabs.harvard.edu/abs/2020GCN.28564....1H} {28564, 1}

\bibitem[\protect\citeauthoryear{{Hennessy} et~al.,}{{Hennessy} et~al.}{2023}]{hennessy2023}
{Hennessy} A.,  et~al., 2023, \mn@doi [\mnras] {10.1093/mnras/stad2670}, \href {https://ui.adsabs.harvard.edu/abs/2023MNRAS.526..106H} {526, 106}

\bibitem[\protect\citeauthoryear{{James}, {Prochaska}, {Macquart}, {North-Hickey}, {Bannister}  \& {Dunning}}{{James} et~al.}{2022}]{james2022}
{James} C.~W.,  {Prochaska} J.~X.,  {Macquart} J.~P.,  {North-Hickey} F.~O.,  {Bannister} K.~W.,   {Dunning} A.,  2022, \mn@doi [\mnras] {10.1093/mnras/stab3051}, \href {https://ui.adsabs.harvard.edu/abs/2022MNRAS.509.4775J} {509, 4775}

\bibitem[\protect\citeauthoryear{{Jia}, {Hu}, {Yang}, {Zhang}  \& {Wang}}{{Jia} et~al.}{2022}]{Jia22}
{Jia} X.~D.,  {Hu} J.~P.,  {Yang} J.,  {Zhang} B.~B.,   {Wang} F.~Y.,  2022, \mn@doi [\mnras] {10.1093/mnras/stac2356}, \href {https://ui.adsabs.harvard.edu/abs/2022MNRAS.516.2575J} {516, 2575}

\bibitem[\protect\citeauthoryear{{Jordana-Mitjans} et~al.,}{{Jordana-Mitjans} et~al.}{2022}]{jordana2022}
{Jordana-Mitjans} N.,  et~al., 2022, \mn@doi [\apj] {10.3847/1538-4357/ac972b}, \href {https://ui.adsabs.harvard.edu/abs/2022ApJ...939..106J} {939, 106}

\bibitem[\protect\citeauthoryear{{Kaplan} et~al.,}{{Kaplan} et~al.}{2015}]{kaplan2015}
{Kaplan} D.~L.,  et~al., 2015, \mn@doi [\apjl] {10.1088/2041-8205/814/2/L25}, \href {https://ui.adsabs.harvard.edu/abs/2015ApJ...814L..25K} {814, L25}

\bibitem[\protect\citeauthoryear{{Karuppusamy}, {Stappers}  \& {van Straten}}{{Karuppusamy} et~al.}{2010}]{karuppusamy2010}
{Karuppusamy} R.,  {Stappers} B.~W.,   {van Straten} W.,  2010, \mn@doi [\aap] {10.1051/0004-6361/200913729}, \href {https://ui.adsabs.harvard.edu/abs/2010A&A...515A..36K} {515, A36}

\bibitem[\protect\citeauthoryear{{Kasen} \& {Bildsten}}{{Kasen} \& {Bildsten}}{2010}]{kasen2010}
{Kasen} D.,  {Bildsten} L.,  2010, \mn@doi [\apj] {10.1088/0004-637X/717/1/245}, \href {https://ui.adsabs.harvard.edu/abs/2010ApJ...717..245K} {717, 245}

\bibitem[\protect\citeauthoryear{{Kirsten}, {Snelders}, {Jenkins}, {Nimmo}, {van den Eijnden}, {Hessels}, {Gawro{\'n}ski}  \& {Yang}}{{Kirsten} et~al.}{2021}]{kirsten2021}
{Kirsten} F.,  {Snelders} M.~P.,  {Jenkins} M.,  {Nimmo} K.,  {van den Eijnden} J.,  {Hessels} J.~W.~T.,  {Gawro{\'n}ski} M.~P.,   {Yang} J.,  2021, \mn@doi [Nature Astronomy] {10.1038/s41550-020-01246-3}, \href {https://ui.adsabs.harvard.edu/abs/2021NatAs...5..414K} {5, 414}

\bibitem[\protect\citeauthoryear{{Kouveliotou}, {Meegan}, {Fishman}, {Bhat}, {Briggs}, {Koshut}, {Paciesas}  \& {Pendleton}}{{Kouveliotou} et~al.}{1993}]{Kouveliotou93}
{Kouveliotou} C.,  {Meegan} C.~A.,  {Fishman} G.~J.,  {Bhat} N.~P.,  {Briggs} M.~S.,  {Koshut} T.~M.,  {Paciesas} W.~S.,   {Pendleton} G.~N.,  1993, \mn@doi [\apjl] {10.1086/186969}, \href {https://ui.adsabs.harvard.edu/abs/1993ApJ...413L.101K} {413, L101}

\bibitem[\protect\citeauthoryear{{Lasky}, {Haskell}, {Ravi}, {Howell}  \& {Coward}}{{Lasky} et~al.}{2014}]{lasky2014}
{Lasky} P.~D.,  {Haskell} B.,  {Ravi} V.,  {Howell} E.~J.,   {Coward} D.~M.,  2014, \mn@doi [Phys. Rev. D.] {10.1103/PhysRevD.89.047302}, \href {https://ui.adsabs.harvard.edu/abs/2014PhRvD..89d7302L} {89, 047302}

\bibitem[\protect\citeauthoryear{{Lien} et~al.,}{{Lien} et~al.}{2016}]{Lien16}
{Lien} A.,  et~al., 2016, \mn@doi [\apj] {10.3847/0004-637X/829/1/7}, \href {https://ui.adsabs.harvard.edu/abs/2016ApJ...829....7L} {829, 7}

\bibitem[\protect\citeauthoryear{{Loi} et~al.,}{{Loi} et~al.}{2015}]{loi2015}
{Loi} S.~T.,  et~al., 2015, \mn@doi [\mnras] {10.1093/mnras/stv1808}, \href {https://ui.adsabs.harvard.edu/abs/2015MNRAS.453.2731L} {453, 2731}

\bibitem[\protect\citeauthoryear{{L{\"u}}, {Zhang}, {Lei}, {Li}  \& {Lasky}}{{L{\"u}} et~al.}{2015}]{lu2015}
{L{\"u}} H.-J.,  {Zhang} B.,  {Lei} W.-H.,  {Li} Y.,   {Lasky} P.~D.,  2015, \mn@doi [\apj] {10.1088/0004-637X/805/2/89}, \href {https://ui.adsabs.harvard.edu/abs/2015ApJ...805...89L} {805, 89}

\bibitem[\protect\citeauthoryear{{Lyubarsky}}{{Lyubarsky}}{2014}]{lyubarsky2014}
{Lyubarsky} Y.,  2014, \mn@doi [\mnras] {10.1093/mnrasl/slu046}, \href {https://ui.adsabs.harvard.edu/abs/2014MNRAS.442L...9L} {442, L9}

\bibitem[\protect\citeauthoryear{{Macquart} et~al.,}{{Macquart} et~al.}{2020}]{macquart2020}
{Macquart} J.~P.,  et~al., 2020, \mn@doi [\nat] {10.1038/s41586-020-2300-2}, \href {https://ui.adsabs.harvard.edu/abs/2020Natur.581..391M} {581, 391}

\bibitem[\protect\citeauthoryear{{Metzger}, {Giannios}, {Thompson}, {Bucciantini}  \& {Quataert}}{{Metzger} et~al.}{2011}]{metzger2011}
{Metzger} B.~D.,  {Giannios} D.,  {Thompson} T.~A.,  {Bucciantini} N.,   {Quataert} E.,  2011, \mn@doi [\mnras] {10.1111/j.1365-2966.2011.18280.x}, \href {https://ui.adsabs.harvard.edu/abs/2011MNRAS.413.2031M} {413, 2031}

\bibitem[\protect\citeauthoryear{{Metzger}, {Thompson}  \& {Quataert}}{{Metzger} et~al.}{2018}]{metzger2018}
{Metzger} B.~D.,  {Thompson} T.~A.,   {Quataert} E.,  2018, \mn@doi [\apj] {10.3847/1538-4357/aab095}, \href {https://ui.adsabs.harvard.edu/abs/2018ApJ...856..101M} {856, 101}

\bibitem[\protect\citeauthoryear{{Moroianu}, {Wen}, {James}, {Ai}, {Kovalam}, {Panther}  \& {Zhang}}{{Moroianu} et~al.}{2022}]{moroianu2022}
{Moroianu} A.,  {Wen} L.,  {James} C.~W.,  {Ai} S.,  {Kovalam} M.,  {Panther} F.,   {Zhang} B.,  2022, \mn@doi [arXiv e-prints] {10.48550/arXiv.2212.00201}, \href {https://ui.adsabs.harvard.edu/abs/2022arXiv221200201M} {p. arXiv:2212.00201}

\bibitem[\protect\citeauthoryear{{Nimmo} et~al.,}{{Nimmo} et~al.}{2022}]{nimmo2022}
{Nimmo} K.,  et~al., 2022, \mn@doi [Nature Astronomy] {10.1038/s41550-021-01569-9}, \href {https://ui.adsabs.harvard.edu/abs/2022NatAs...6..393N} {6, 393}

\bibitem[\protect\citeauthoryear{{Offringa}, {de Bruyn}, {Biehl}, {Zaroubi}, {Bernardi}  \& {Pandey}}{{Offringa} et~al.}{2010}]{offringa2010}
{Offringa} A.~R.,  {de Bruyn} A.~G.,  {Biehl} M.,  {Zaroubi} S.,  {Bernardi} G.,   {Pandey} V.~N.,  2010, \mn@doi [\mnras] {10.1111/j.1365-2966.2010.16471.x}, \href {https://ui.adsabs.harvard.edu/abs/2010MNRAS.405..155O} {405, 155}

\bibitem[\protect\citeauthoryear{{Offringa}, {van de Gronde}  \& {Roerdink}}{{Offringa} et~al.}{2012}]{offringa2012}
{Offringa} A.~R.,  {van de Gronde} J.~J.,   {Roerdink} J.~B.~T.~M.,  2012, \mn@doi [\aap] {10.1051/0004-6361/201118497}, \href {https://ui.adsabs.harvard.edu/abs/2012A&A...539A..95O} {539, A95}

\bibitem[\protect\citeauthoryear{{Offringa} et~al.,}{{Offringa} et~al.}{2014}]{offringa2014}
{Offringa} A.~R.,  et~al., 2014, \mn@doi [\mnras] {10.1093/mnras/stu1368}, \href {https://ui.adsabs.harvard.edu/abs/2014MNRAS.444..606O} {444, 606}

\bibitem[\protect\citeauthoryear{{{\"O}zel}, {Psaltis}, {Ransom}, {Demorest}  \& {Alford}}{{{\"O}zel} et~al.}{2010}]{ozel2010}
{{\"O}zel} F.,  {Psaltis} D.,  {Ransom} S.,  {Demorest} P.,   {Alford} M.,  2010, \mn@doi [\apjl] {10.1088/2041-8205/724/2/L199}, \href {https://ui.adsabs.harvard.edu/abs/2010ApJ...724L.199O} {724, L199}

\bibitem[\protect\citeauthoryear{{Petroff}, {Hessels}  \& {Lorimer}}{{Petroff} et~al.}{2022}]{petroff2022}
{Petroff} E.,  {Hessels} J.~W.~T.,   {Lorimer} D.~R.,  2022, \mn@doi [Astron. \& Astrophys. Rev.] {10.1007/s00159-022-00139-w}, \href {https://ui.adsabs.harvard.edu/abs/2022A&ARv..30....2P} {30, 2}

\bibitem[\protect\citeauthoryear{{Pietka}, {Fender}  \& {Keane}}{{Pietka} et~al.}{2015}]{pietka2015}
{Pietka} M.,  {Fender} R.~P.,   {Keane} E.~F.,  2015, \mn@doi [\mnras] {10.1093/mnras/stu2335}, \href {https://ui.adsabs.harvard.edu/abs/2015MNRAS.446.3687P} {446, 3687}

\bibitem[\protect\citeauthoryear{Prochaska, Simha, Law, Tejos  \& mneeleman}{Prochaska et~al.}{2019}]{prochaska2019}
Prochaska J.~X.,  Simha S.,  Law C.,  Tejos N.,   mneeleman 2019, FRBs/FRB: First DOI release of this repository, \mn@doi{10.5281/zenodo.3403651}, \url {https://doi.org/10.5281/zenodo.3403651}

\bibitem[\protect\citeauthoryear{{Rajwade}}{{Rajwade}}{prep}]{rajwade}
{Rajwade} K.,  in prep

\bibitem[\protect\citeauthoryear{{Ravi} \& {Lasky}}{{Ravi} \& {Lasky}}{2014}]{ravi2014}
{Ravi} V.,  {Lasky} P.~D.,  2014, \mn@doi [\mnras] {10.1093/mnras/stu720}, \href {https://ui.adsabs.harvard.edu/abs/2014MNRAS.441.2433R} {441, 2433}

\bibitem[\protect\citeauthoryear{{Rezzolla}, {Giacomazzo}, {Baiotti}, {Granot}, {Kouveliotou}  \& {Aloy}}{{Rezzolla} et~al.}{2011}]{rezzolla2011}
{Rezzolla} L.,  {Giacomazzo} B.,  {Baiotti} L.,  {Granot} J.,  {Kouveliotou} C.,   {Aloy} M.~A.,  2011, \mn@doi [\apjl] {10.1088/2041-8205/732/1/L6}, \href {https://ui.adsabs.harvard.edu/abs/2011ApJ...732L...6R} {732, L6}

\bibitem[\protect\citeauthoryear{{Rosswog}}{{Rosswog}}{2007}]{rosswog2007}
{Rosswog} S.,  2007, \mn@doi [\mnras] {10.1111/j.1745-3933.2007.00284.x}, \href {https://ui.adsabs.harvard.edu/abs/2007MNRAS.376L..48R} {376, L48}

\bibitem[\protect\citeauthoryear{{Rowlinson} \& {Anderson}}{{Rowlinson} \& {Anderson}}{2019}]{rowlinson2019b}
{Rowlinson} A.,  {Anderson} G.~E.,  2019, \mn@doi [\mnras] {10.1093/mnras/stz2295}, \href {https://ui.adsabs.harvard.edu/abs/2019MNRAS.489.3316R} {489, 3316}

\bibitem[\protect\citeauthoryear{{Rowlinson}, {O'Brien}, {Metzger}, {Tanvir}  \& {Levan}}{{Rowlinson} et~al.}{2013}]{rowlinson2013}
{Rowlinson} A.,  {O'Brien} P.~T.,  {Metzger} B.~D.,  {Tanvir} N.~R.,   {Levan} A.~J.,  2013, \mn@doi [\mnras] {10.1093/mnras/sts683}, \href {http://adsabs.harvard.edu/abs/2013MNRAS.430.1061R} {430, 1061}

\bibitem[\protect\citeauthoryear{{Rowlinson}, {Gompertz}, {Dainotti}, {O'Brien}, {Wijers}  \& {van der Horst}}{{Rowlinson} et~al.}{2014}]{rowlinson2014}
{Rowlinson} A.,  {Gompertz} B.~P.,  {Dainotti} M.,  {O'Brien} P.~T.,  {Wijers} R.~A.~M.~J.,   {van der Horst} A.~J.,  2014, \mn@doi [\mnras] {10.1093/mnras/stu1277}, \href {https://ui.adsabs.harvard.edu/abs/2014MNRAS.443.1779R} {443, 1779}

\bibitem[\protect\citeauthoryear{{Rowlinson} et~al.,}{{Rowlinson} et~al.}{2016}]{rowlinson2016}
{Rowlinson} A.,  et~al., 2016, \mn@doi [\mnras] {10.1093/mnras/stw451}, \href {https://ui.adsabs.harvard.edu/abs/2016MNRAS.458.3506R} {458, 3506}

\bibitem[\protect\citeauthoryear{{Rowlinson}, {Patruno}  \& {O'Brien}}{{Rowlinson} et~al.}{2017}]{rowlinson2017}
{Rowlinson} A.,  {Patruno} A.,   {O'Brien} P.~T.,  2017, \mn@doi [\mnras] {10.1093/mnras/stx2023}, \href {https://ui.adsabs.harvard.edu/abs/2017MNRAS.472.1152R} {472, 1152}

\bibitem[\protect\citeauthoryear{{Rowlinson} et~al.,}{{Rowlinson} et~al.}{2019}]{rowlinson2019}
{Rowlinson} A.,  et~al., 2019, \mn@doi [\mnras] {10.1093/mnras/stz2866}, \href {https://ui.adsabs.harvard.edu/abs/2019MNRAS.490.3483R} {490, 3483}

\bibitem[\protect\citeauthoryear{{Rowlinson} et~al.,}{{Rowlinson} et~al.}{2021}]{rowlinson2021}
{Rowlinson} A.,  et~al., 2021, \mn@doi [\mnras] {10.1093/mnras/stab2060}, \href {https://ui.adsabs.harvard.edu/abs/2021MNRAS.506.5268R} {506, 5268}

\bibitem[\protect\citeauthoryear{{Rowlinson} et~al.,}{{Rowlinson} et~al.}{2022}]{rowlinson2022}
{Rowlinson} A.,  et~al., 2022, \mn@doi [\mnras] {10.1093/mnras/stac2460}, \href {https://ui.adsabs.harvard.edu/abs/2022MNRAS.517.2894R} {517, 2894}

\bibitem[\protect\citeauthoryear{Ruhe, Kuiack, Rowlinson, Wijers  \& Forr{\'e}}{Ruhe et~al.}{2022}]{ruhe2022detecting}
Ruhe D.,  Kuiack M.,  Rowlinson A.,  Wijers R.,   Forr{\'e} P.,  2022, Astronomy and Computing, 38, 100512

\bibitem[\protect\citeauthoryear{{Schlafly} \& {Finkbeiner}}{{Schlafly} \& {Finkbeiner}}{2011}]{Schlafly}
{Schlafly} E.~F.,  {Finkbeiner} D.~P.,  2011, \mn@doi [\apj] {10.1088/0004-637X/737/2/103}, \href {https://ui.adsabs.harvard.edu/abs/2011ApJ...737..103S} {737, 103}

\bibitem[\protect\citeauthoryear{{Spitler} et~al.,}{{Spitler} et~al.}{2016}]{spitler2016}
{Spitler} L.~G.,  et~al., 2016, \mn@doi [\nat] {10.1038/nature17168}, \href {https://ui.adsabs.harvard.edu/abs/2016Natur.531..202S} {531, 202}

\bibitem[\protect\citeauthoryear{{Starling}, {Rowlinson}, {van der Horst}  \& {Wijers}}{{Starling} et~al.}{2020}]{starling2020}
{Starling} R.~L.~C.,  {Rowlinson} A.,  {van der Horst} A.~J.,   {Wijers} R.~A.~M.~J.,  2020, \mn@doi [\mnras] {10.1093/mnras/staa1168}, \href {https://ui.adsabs.harvard.edu/abs/2020MNRAS.494.5787S} {494, 5787}

\bibitem[\protect\citeauthoryear{{Suvorov} \& {Glampedakis}}{{Suvorov} \& {Glampedakis}}{2022}]{suvorov2022}
{Suvorov} A.~G.,  {Glampedakis} K.,  2022, \mn@doi [\prd] {10.1103/PhysRevD.105.L061302}, \href {https://ui.adsabs.harvard.edu/abs/2022PhRvD.105f1302S} {105, L061302}

\bibitem[\protect\citeauthoryear{{Swinbank} et~al.,}{{Swinbank} et~al.}{2015}]{swinbank2015}
{Swinbank} J.~D.,  et~al., 2015, \mn@doi [Astronomy and Computing] {10.1016/j.ascom.2015.03.002}, \href {https://ui.adsabs.harvard.edu/abs/2015A&C....11...25S} {11, 25}

\bibitem[\protect\citeauthoryear{{Taylor}, {Manchester}  \& {Lyne}}{{Taylor} et~al.}{1993}]{taylor1993}
{Taylor} J.~H.,  {Manchester} R.~N.,   {Lyne} A.~G.,  1993, \mn@doi [\apjs] {10.1086/191832}, \href {https://ui.adsabs.harvard.edu/abs/1993ApJS...88..529T} {88, 529}

\bibitem[\protect\citeauthoryear{{Tian} et~al.,}{{Tian} et~al.}{2022}]{tian2022}
{Tian} J.,  et~al., 2022, \mn@doi [\pasa] {10.1017/pasa.2021.58}, \href {https://ui.adsabs.harvard.edu/abs/2022PASA...39....3T} {39, e003}

\bibitem[\protect\citeauthoryear{{Tingay} et~al.,}{{Tingay} et~al.}{2013}]{tingay2013}
{Tingay} S.~J.,  et~al., 2013, \mn@doi [\pasa] {10.1017/pasa.2012.007}, \href {https://ui.adsabs.harvard.edu/abs/2013PASA...30....7T} {30, e007}

\bibitem[\protect\citeauthoryear{{Totani}}{{Totani}}{2013}]{totani2013}
{Totani} T.,  2013, \mn@doi [\pasj] {10.1093/pasj/65.5.L12}, \href {https://ui.adsabs.harvard.edu/abs/2013PASJ...65L..12T} {65, L12}

\bibitem[\protect\citeauthoryear{{Usov} \& {Katz}}{{Usov} \& {Katz}}{2000}]{usov2000}
{Usov} V.~V.,  {Katz} J.~I.,  2000, \mn@doi [\aap] {10.48550/arXiv.astro-ph/0002278}, \href {https://ui.adsabs.harvard.edu/abs/2000A&A...364..655U} {364, 655}

\bibitem[\protect\citeauthoryear{{Williams} et~al.,}{{Williams} et~al.}{2016}]{williams2016}
{Williams} W.~L.,  et~al., 2016, \mn@doi [\mnras] {10.1093/mnras/stw1056}, \href {https://ui.adsabs.harvard.edu/abs/2016MNRAS.460.2385W} {460, 2385}

\bibitem[\protect\citeauthoryear{{Yamasaki} \& {Totani}}{{Yamasaki} \& {Totani}}{2020}]{yamasaki2020}
{Yamasaki} S.,  {Totani} T.,  2020, \mn@doi [\apj] {10.3847/1538-4357/ab58c4}, \href {https://ui.adsabs.harvard.edu/abs/2020ApJ...888..105Y} {888, 105}

\bibitem[\protect\citeauthoryear{{Yao}, {Manchester}  \& {Wang}}{{Yao} et~al.}{2017}]{yao2017}
{Yao} J.~M.,  {Manchester} R.~N.,   {Wang} N.,  2017, \mn@doi [\apj] {10.3847/1538-4357/835/1/29}, \href {https://ui.adsabs.harvard.edu/abs/2017ApJ...835...29Y} {835, 29}

\bibitem[\protect\citeauthoryear{{Yao}, {Manchester}  \& {Wang}}{{Yao} et~al.}{2019}]{yao2019}
{Yao} J.,  {Manchester} R.~N.,   {Wang} N.,  2019, {YMW16: Electron-density model}, Astrophysics Source Code Library, record ascl:1908.022 (\mn@eprint {ascl} {1908.022})

\bibitem[\protect\citeauthoryear{{Zhang}}{{Zhang}}{2014}]{zhang2014}
{Zhang} B.,  2014, \mn@doi [\apjl] {10.1088/2041-8205/780/2/L21}, \href {http://adsabs.harvard.edu/abs/2014ApJ...780L..21Z} {780, L21}

\bibitem[\protect\citeauthoryear{{Zhang} \& {M{\'e}sz{\'a}ros}}{{Zhang} \& {M{\'e}sz{\'a}ros}}{2001}]{zhang2001}
{Zhang} B.,  {M{\'e}sz{\'a}ros} P.,  2001, \mn@doi [\apjl] {10.1086/320255}, \href {http://adsabs.harvard.edu/abs/2001ApJ...552L..35Z} {552, L35}

\bibitem[\protect\citeauthoryear{{de Gasperin} et~al.,}{{de Gasperin} et~al.}{2019}]{degasperin2019}
{de Gasperin} F.,  et~al., 2019, \mn@doi [\aap] {10.1051/0004-6361/201833867}, \href {https://ui.adsabs.harvard.edu/abs/2019A&A...622A...5D} {622, A5}

\bibitem[\protect\citeauthoryear{{de Ruiter}, {Meyers}, {Rowlinson}, {Shimwell}, {Ruhe}  \& {Wijers}}{{de Ruiter} et~al.}{2023}]{deruiter}
{de Ruiter} I.,  {Meyers} Z.~S.,  {Rowlinson} A.,  {Shimwell} T.~W.,  {Ruhe} D.,   {Wijers} R. A.~M.~J.,  2023, \mn@doi [arXiv e-prints] {10.48550/arXiv.2311.07394}, \href {https://ui.adsabs.harvard.edu/abs/2023arXiv231107394D} {p. arXiv:2311.07394}

\bibitem[\protect\citeauthoryear{{van Haarlem} et~al.,}{{van Haarlem} et~al.}{2013}]{vanhaarlem2013}
{van Haarlem} M.~P.,  et~al., 2013, \mn@doi [\aap] {10.1051/0004-6361/201220873}, \href {https://ui.adsabs.harvard.edu/abs/2013A&A...556A...2V} {556, A2}

\bibitem[\protect\citeauthoryear{{van Weeren} et~al.,}{{van Weeren} et~al.}{2016}]{vanweeren2016}
{van Weeren} R.~J.,  et~al., 2016, \mn@doi [\apjs] {10.3847/0067-0049/223/1/2}, \href {https://ui.adsabs.harvard.edu/abs/2016ApJS..223....2V} {223, 2}

\makeatother
\end{thebibliography}



\bsp	
\label{lastpage}
\end{document}